\documentclass[aps,pra,twocolumn,showpacs,groupedaddress,superscriptaddress,nofootinbib,floatfix,preprintnumbers,longbibliography]{revtex4-2}

\usepackage{amssymb,amsmath,graphicx,color,qcircuit}
\usepackage{bm}
\usepackage[tight]{subfigure}
\usepackage[export]{adjustbox}
\usepackage{braket,mathtools}
\usepackage[colorlinks=true,citecolor=blue,linkcolor=red]{hyperref}
\usepackage[capitalise,compress]{cleveref}
\usepackage[english]{babel}
\usepackage{float}
\usepackage{tabularx}
\usepackage{booktabs}
\usepackage[utf8]{inputenc}
\usepackage[makeroom]{cancel}
\makeatletter
\let\ORIbbl@fixname\bbl@fixname
\def\bbl@fixname#1{%
  \@ifundefined{languagealias@\expandafter\string#1}
    {\ORIbbl@fixname#1}
    {\edef\languagename{\@nameuse{languagealias@#1}}}%
}
\newcommand{\definelanguagealias}[2]{%
  \@namedef{languagealias@#1}{#2}%
}
\makeatother

\definelanguagealias{en}{english}
\definelanguagealias{EN}{english}
\newcommand{\cmt}[1]{} 

\newcommand{\comm}[1]{\left[#1\right]}

\newcommand{\abs}[1]{\left | #1 \right|}
\newcommand{\norm}[1]{\left \Vert #1 \right\Vert}
\renewcommand{\O}[1]{\mathcal{O}\left(#1\right)}
\newcolumntype{C}{>{$}c<{$}}
\newcolumntype{R}{>{$}r<{$}}
\newcolumntype{L}{>{$}l<{$}}
\newcolumntype{Y}{>{\centering\arraybackslash}X}

\begin{document}

\title{
Digital Quantum Simulation of the Schwinger Model and 
\\Testing Symmetry Protection with Trapped Ions
}

\author{Nhung H. Nguyen}
\affiliation{
Joint Quantum Institute and Department of Physics,
University of Maryland, College Park, Maryland 20742, USA}

\author{Minh C. Tran}
\affiliation{Center for Theoretical Physics, Massachusetts Institute of Technology, Cambridge, Massachusetts 02139, USA}
\affiliation{Department of Physics, Harvard University, Cambridge, Massachusetts 02138, USA}

\author{Yingyue Zhu}
\affiliation{
Joint Quantum Institute and Department of Physics,
University of Maryland, College Park, Maryland 20742, USA}

\author{Alaina~M.~Green}
\affiliation{
Joint Quantum Institute and Department of Physics,
University of Maryland, College Park, Maryland 20742, USA}

\author{C.~Huerta~Alderete}
\affiliation{
Joint Quantum Institute and Department of Physics,
University of Maryland, College Park, Maryland 20742, USA}

\author{Zohreh Davoudi}
\affiliation{
Maryland Center for Fundamental Physics and Department of Physics, 
University of Maryland, College Park, Maryland 20742, USA}

\author{Norbert M. Linke}
\affiliation{
Joint Quantum Institute and Department of Physics,
University of Maryland, College Park, Maryland 20742, USA}

\date{\today}

\preprint{MIT-CTP/5384, UMD-PP-021-08
}

\begin{abstract}
Tracking the dynamics of physical systems in real time is a prime application of quantum computers. 
Using a trapped-ion system with up to six qubits, we simulate the real-time dynamics of a lattice gauge theory in 1+1 dimensions, i.e., the lattice Schwinger model, and demonstrate non-perturbative effects such as pair creation for times much longer than previously accessible. We study the gate requirement of two formulations of the model using the Suzuki-Trotter product formula, as well as the trade-off between errors from the ordering of the Hamiltonian terms, the Trotter step size, and experimental imperfections. To mitigate experimental errors, a recent symmetry-protection protocol for suppressing coherent errors and a symmetry-inspired post-selection scheme are applied.
This work demonstrates the integrated theoretical, algorithmic, and experimental approach that is essential for efficient simulation of lattice gauge theories and other complex physical systems.

\end{abstract}

\maketitle

\section{Introduction
\label{sec:intro}}
\noindent
An exciting prospect for quantum computers is the simulation of complex physical systems~\cite{Feynman:1982, lloydUniversalQuantumSimulators1996}. Digital quantum computers can simulate a wide range of physical Hamiltonians since one can often find efficient circuit decompositions to approximate their dynamics.  However, the number of qubits and the achievable circuit depth in experiments are limited. 
In particular, it is challenging to simulate larger systems for longer time while maintaining the fidelity of the simulations. 
To address this challenge, there has been substantial theoretical research on finding optimal simulation algorithms with improved theoretical error bounds~\cite{childsTheoryTrotterError2020, childs2012hamiltonian, berry2015simulating, low2017optimal, low2019hamiltonian, chakraborty2018power, gilyen2019quantum, kalev2021quantum} and better empirical performance~\cite{childsNearlyOptimalLattice2019, stetina2020simulating}, as well as efficient circuit decomposition with resource analysis for a range of problems~\cite{PRXQuantum.2.030305,PhysRevResearch.3.033055,Huggins2021,PhysRevX.8.041015,PhysRevX.8.011044,stetina2020simulating, jordan2012quantum, Klco:2018zqz, Barata:2020jtq, Lamm:2019bik, shawQuantumAlgorithmsSimulating2020b, Ciavarella:2021nmj, kan2021lattice, kreshchuk2020quantum, liu2020quantum, Paulson:2020zjd, Bauer:2019qxa}. Furthermore, symmetry-protection schemes have emerged to suppress algorithmic and experimental errors~\cite{tranFasterDigitalQuantum2020, Lamm:2020jwv, halimehGaugeSymmetryProtectionUsing2020a, Halimeh:2021vzf, Stannigel:2013zka, kasperNonAbelianGaugeInvariance2020, Stryker:2018efp, Raychowdhury:2018osk}, along with noise-mitigation schemes for extrapolating to the noiseless limit of observables~\cite{kandala2018extending, temme2017error, endo2018practical, tannu2019mitigating, wallman2016noise, shaw2021classical, he2020zero, czarnik2021error, vovrosh2021simple}.
As a result, it is of great interest to study how well these recent algorithmic advances, which mostly concern asymptotic regimes with a large number of qubits and long evolution times, apply in a current experiment.

The physical system we consider is a low-dimensional lattice gauge theory. Gauge field theories are the underlying formalism describing interactions among elementary particles in the Standard Model, and are prime candidates for modeling physics beyond the Standard Model. They further provide a powerful theoretical framework for describing low-energy excitations in condensed-matter systems. There has been tremendous progress in applying non-perturbative methods to solve lattice gauge theories in various systems and coupling regimes~\cite{Aoki:2021kgd, Beane:2010em, Ding:2015ona, Briceno:2017max, Ratti:2018ksb, Davoudi:2020ngi}. However, evaluating the real-time dynamics of gauge field theories remains challenging in the strong-coupling regime, where a notorious sign problem halts Monte-Carlo-based classical simulations~\cite{Calzetta:2008iqa}. Both fermionic and bosonic degrees of freedom in gauge theories, when defined on a discretized spacetime, can be mapped to spins and, in principle, efficiently simulated using quantum simulators. A simple Abelian low-dimensional gauge theory is quantum electrodynamics in 1+1 dimensions, or the Schwinger model \cite{schwingerGaugeInvarianceVacuum1951}. It exhibits non-trivial dynamics similar to those seen in quantum chromodynamics in 3+1 dimensions, i.e., the theory of the strong force in nature, including particle-antiparticle pair creation, chiral symmetry breaking, confinement, and a non-trivial $\theta$-vacuum~\cite{schwingerGaugeInvarianceMass1962, PhysRev.91.713}. The Hamiltonian formulation of the lattice Schwinger model, i.e., the Schwinger model defined in a discretized space and with continuous time, has served as a testbed for numerous computational techniques in recent years, including quantum simulation and computation. In particular, there have been several theoretical proposals for analog quantum simulation~\cite{banerjeeAtomicQuantumSimulation2012a, zoharQuantumSimulationsLattice2015, yangAnalogQuantumSimulation2016, davoudiAnalogQuantumSimulations2020, suraceLatticeGaugeTheories2020, luo2020framework, andrade2021engineering} and gate-based quantum algorithms~\cite{muschikWilsonLatticeGauge2017, klcoQuantumClassicalComputationSchwinger2018a, lu2019simulations, Chakraborty:2020uhf, shawQuantumAlgorithmsSimulating2020b, stryker2021shearing, kan2021lattice, davoudiSimulatingQuantumField2021, Pederiva:2021tcd, Rajput:2021khs} of the Schwinger model, along with experimental implementations on various quantum platforms such as trapped ions~\cite{martinezRealtimeDynamicsLattice2016,kokailSelfverifyingVariationalQuantum2019}, Rydberg atoms~\cite{bernienProbingManybodyDynamics2017a,suraceLatticeGaugeTheories2020}, ultracold atoms~\cite{milScalableRealizationLocal2020, zhou2021thermalization}, and superconducting qubits~\cite{klcoQuantumClassicalComputationSchwinger2018a, de2021quantum}. 

Inspired by the first digital simulation of the Schwinger model in Ref.~\cite{martinezRealtimeDynamicsLattice2016}, we revisit the simulation to answer the following questions: i) Are larger and longer simulations of lattice gauge theories viable on present hardware? ii) Given recent progress in quantifying the theoretical and empirical bounds on simulation algorithms, what are the resource requirements of simulating the lattice Schwinger model in the purely fermionic formulation with long-range interactions versus the fermion-boson formulation with only local interactions? iii) What are the theoretical considerations in decomposing the time evolution operator using product formulas, e.g., how does the Trotter error depend on the ordering of terms in each Trotter step of the evolution, what order of the product formula should one use, and how small should the Trotter steps be, considering the anticipated size of the experimental error? iv) Can recent cost-efficient symmetry protection protocols provide insights into the nature of experimental errors?

To address these questions, we realize experimentally the time dynamics of the lattice Schwinger model in its staggered formulation~\cite{kogutHamiltonianFormulationWilson1975a}, and within its purely fermionic representation, for two-, four-, and six-site theories. Along with the experimental demonstration, different term orderings and product formulas are studied, and the gate complexity of the algorithm used here is compared with another algorithm in the local formulation of the same theory~\cite{shawQuantumAlgorithmsSimulating2020b}. The symmetry protection of Ref.~\cite{tranFasterDigitalQuantum2020} is implemented for the first time
in experiment, in addition to a simple symmetry-inspired post-selection.

\section{Theory and algorithm considerations
\label{sec:theory}}
\noindent
In the staggered formulation of the lattice Schwinger model introduced by Kogut and Susskind \cite{kogutHamiltonianFormulationWilson1975a}, the two-component matter field at one spatial site is split into two one-component fields, $\hat{\psi}$, each occupying one site of the staggered lattice. This staggering corresponds to placing electrons at odd sites and positrons at even sites. The electric field, $\hat{E}$, and the corresponding gauge-link variable, $\hat{U}$, are defined on the link connecting the two adjacent staggered sites. The Hamiltonian in natural units is:
\begin{eqnarray}
&&\hat{H}_{\rm KS}=\frac{i}{2a}\sum_{n=1}^{N-1}(\hat{\psi}^\dagger_n\hat{U}_n\hat{\psi}_{n+1}-\text{H.c.})+\frac{g^2a}{2}\sum_{n=1}^{N-1}\hat{E}^2_n+
\nonumber
\\
&&\hspace{4 cm}
\frac{m}2\sum_{n=1}^N(-1)^n\hat{\psi}^\dagger_n\hat{\psi}_n,
\label{eq:Horiginal}
\end{eqnarray}
where $N$ is the number of staggered sites, $m$ is the mass of the fermions, $a$ denotes the lattice spacing, and $g$ is the coupling constant. The first term in the Hamiltonian involves gauge-boson-assisted fermionic hopping between nearest-neighbor sites. The second term is the energy stored in the electric field. The last term represents the rest mass of the fermions in the staggered formulation. The theory exhibits a local gauge symmetry, which leads to a Gauss's law constraint on the allowed physical states $\ket{\phi}_{\rm phys}$. This constraint enforces the net electric field at each site to be balanced by the electric charge present at the site, i.e., $\hat{G}_n\ket{\phi}_{\rm phys}=0$ with $\hat{G}_n=\hat{E}_n-\hat{E}_{n-1}-\hat{\psi}^\dagger_n\hat{\psi}_n+\frac{1}{2}[1-(-1)^n]$ for all $n$. 

The matter-field operators in Eq.~\eqref{eq:Horiginal} can be mapped to spin operators by the Jordan-Wigner transformation: $\hat\psi_n = \prod_{l<n}(i\hat\sigma_l^Z)\hat\sigma_n^-$ and $\hat\psi_n^\dagger = \prod_{l<n}(-i\hat\sigma_l^Z)\hat\sigma_n^+$, with $\hat\sigma_n^\pm=\frac{1}{2}(\hat\sigma_n^X \pm i \hat\sigma_n^Y)$. With open boundary conditions (OBCs), the gauge links and the electric fields can be eliminated upon a gauge transformation and the application of Gauss's law~\cite{hamerSeriesExpansionsMassive1997}. Without loss of generality, the electric field coming into the lattice is set to zero, and the resulting spin Hamiltonian reads
\begin{align}
    \hat{H}
    &=x\sum_{n=1}^{N-1}(\hat{\sigma}^X_n \hat{\sigma}^X_{n+1}+\hat{\sigma}^Y_n \hat{\sigma}^Y_{n+1})+
    \nonumber\\
    & \hspace{0.4 cm} \frac14\sum_{n=1}^{N-1}\Bigg[\sum_{m=1}^n\Big(\hat\sigma_m^Z+(-1)^m\Big)\Bigg]^2+\mu\sum_{n=1}^N(-1)^n\frac{\hat{\sigma}^Z_n+1}{2}
    \nonumber  \\
    & \equiv \hat{H}^x+ \hat{H}^{ZZ}+\hat{H}^{Z}+{\rm const.}
    \label{eq:Hspin}
\end{align}
Here, the original Hamiltonian is rescaled by $\frac{2}{ag^2}$, i.e., $\hat{H}=\frac{2}{ag^2}\hat{H}_{\rm KS}$, so that $\hat{H}$ in Eq.~(\ref{eq:Hspin}) is dimensionless with parameters $x=\frac1{g^2a^2}$ and $\mu=\frac{2m}{ag^2}$. $\hat{H}^x$ denotes the term proportional to the coupling $x$, $\hat{H}^{ZZ}$ consists of (nearly) all-to-all spin-spin interactions proportional to $\hat{\sigma}^Z_n\hat{\sigma}^Z_m$, and $\hat{H}^Z$ is the sum of the terms proportional to $\hat{\sigma}^Z_n$. The constant terms are ignored in the following as they do not affect the evolution. By convention, $\hat\sigma^Z\ket0=\ket0,\hat\sigma^Z\ket1=-\ket1$, and the presence of a particle (antiparticle) is encoded as $\ket1$ ($\ket0$) at an odd (even) site. Therefore, for an $N$-site system, the bare-vacuum state, i.e., the ground state of the theory in the $x=0$ limit, is $\ket{\psi_0}\equiv (\ket{01})^{\otimes N/2}$.

The goal is to study the real-time dynamics of bare-vacuum fluctuations and particle-antiparticle pair creation. One can therefore perform a quench experiment where the simulation is initiated in the bare-vacuum state, $\ket{\psi(0)}=\ket{\psi_0}$, which is then evolved via the Hamiltonian in Eq.~(\ref{eq:Hspin}) with $x\neq0$. The survival probability of the bare-vacuum state is
\begin{align}
    P_{\text{vac}}(t)=
    \abs{\braket{\psi(0)|\psi(t)}}^2
    =\abs{\bra{\psi(0)} e^{-it\hat{H}}\ket{\psi(0)}}^2, \label{eq:Pvac}
\end{align}
and the particle-number density is $\nu(t) =\frac{1}{N}\sum_{n=1}^N\nu_n(t)$ with
\begin{align}
    \nu_n(t)=
    \braket{\psi(t)|\frac{(-1)^n\hat{\sigma}^Z_n+1}{2}|\psi(t)}.
    \label{eq:nu}
\end{align}
In the limit of $N\rightarrow \infty$, the two quantities are related to each other via $\nu(t)=-\frac1N \log(P_\text{vac})$ \cite{schwingerGaugeInvarianceVacuum1951}. The local charge density is defined as 
\begin{align}
    \label{eq:Qn}
    Q_n(t) &\equiv 
    \braket{\psi(t)|\frac{\hat\sigma^Z_n+(-1)^n}{2}|\psi(t)}
    \\
    &=\braket{\psi(t)|\hat{E}_n-\hat{E}_{n-1}|\psi(t)},
\end{align}
where the particle's (antiparticle's) charge is $-1$ ($1$). The local charge density is related to the local particle-number density by $Q_n(t)=(-1)^n\nu_n(t)$. As there is no interaction in the Hamiltonian that changes the total net charge of the system, $\sum_nQ_n$ is conserved. Therefore, the model has a global symmetry operator $\hat S_z=\sum_n\hat\sigma^Z_n$, which is consistent with the symmetry of an XXZ Heisenberg spin Hamiltonian.

To perform the unitary evolution $\hat{\mathcal{U}}(t)=e^{-it\hat{H}}$ with  $\hat{H} \equiv \sum_{k=1}^K\hat{h}_k$ on a quantum computer, with non-commuting Hamiltonian terms $\hat{h}_k$, the evolution can be broken into $r$ smaller time steps of size $\delta t=t/r$. For each time step, the unitary $\hat{\mathcal{U}}(\delta t)$ is approximated by a product formula, which can then be implemented in terms of the available native gates. For the first-order product formula,
\begin{align}
    \hat{\mathcal{S}}_1(\delta t)&= \prod_{k=1}^Ke^{-i\delta t\,\hat{h}_k},
\end{align}
while for the second-order product formula,
\begin{align}
    \hat{\mathcal{S}}_2(\delta t)&=\prod_{k=1}^Ke^{-i\frac{\delta t}2 \hat{h}_k} \prod_{k=K}^1e^{-i\frac{\delta t}2 \hat{h}_k},\label{eq:2nd-product-formula}
\end{align}
where the second product is realized in reversed order. Higher-order formulas $\hat{\mathcal{S}}_{p}(\delta t)$ can also be constructed recursively for even integers $p$~\cite{suzukiGeneralTheoryFractal1991}.

While it is generally difficult to obtain an exact estimate for the Trotter error in the $p$th-order product formula, progress in recent years has resulted in tighter error bounds for them.
For example,  a nearly optimal bound is derived in Ref.~\cite{childsTheoryTrotterError2020}, expressed in terms of the nested commutators between different $\hat h_k$ terms. For the error of the second-order formula:
\begin{align}
    &\norm{\hat{\mathcal{U}}(\delta t) - \hat{\mathcal{S}}_2(\delta t)}
    \nonumber\\
    & \hspace{0.0 cm}\leq \frac{(\delta t)^3}{24} \sum_{k_1 = 1}^{K-1}\norm{\comm{\hat h_{k_1},\comm{\hat h_{k_1},\sum_{k_2 = k_1+1}^K \hat h_{k_2}}}}+\nonumber\\
    &\hspace{0.45 cm} \frac{(\delta t)^3}{12} \sum_{k_1 = 1}^{K-1}\norm{\comm{\sum_{k_3 = k_1+1}^K\hat h_{k_3},\comm{\sum_{k_2 = k_1+1}^K \hat h_{k_2},\hat h_{k_1}}}}.
    \label{eq:exact-comm-bound}
\end{align}
The norms of non-vanishing nested commutators can be further upper-bounded using the triangle inequality $\norm{[A,B]}\leq 2\norm{A}\norm{B}$, resulting in closed-form, albeit looser, error bounds.
Based on this approach, an upper bound for the $p$th-order formulas in simulating a two-body Hamiltonian can be obtained,
\begin{align}
    \norm{\hat{\mathcal{U}}(\delta t) - \hat{\mathcal{S}}_p(\delta t)} 
    \leq \kappa_p  \gamma  \lambda^p(\delta t)^{p+1},
    \label{eq:general-error-bound}
\end{align}
where $\kappa_p = (4\times 5^{p/2-1})^{p+1}$ is a constant, $\lambda$ is the maximum strength of  the sum of the interactions that involve any particular site, and $\gamma = \sum_{k=1}^{K-1}\Vert\hat h_k\Vert$.
Since $\gamma$ only involves the first $K-1$ terms of the Hamiltonian, labeling the index $k$ such that $\hat h_K$ is the term with the largest norm typically results in the smallest error bound.

Given a small $\delta t$, a higher-order formula results in a smaller simulation error.
However, since the gate count of $\hat{\mathcal{S}}_p(\delta t)$ scales as $\O{5^{p/2}}$, increasing $p$ increases the gate complexity of the simulation exponentially.
The optimal choice for $p$ is a balance between the gate depth, the evolution time $t$, and the error tolerance of the simulation.
Since the experimental error can dominate the Trotter error at long evolution times, the first-order product formula, which minimizes the gate count per Trotter step, turns out to be a more suitable choice for the experiment in this work.

\subsection{Term ordering
\label{sec:term-order}}
Generally, the Hamiltonian terms in Eq.~(\ref{eq:Hspin}) do not commute. Therefore, one needs to find the optimal ordering in the application of the product formula to minimize the Trotter error. There are two typical orderings of the terms in a nearest-neighbor Heisenberg spin model. The odd-even ordering is defined as having the odd-leading two-spin terms, e.g., $\hat\sigma^\tau_{2k-1}\hat\sigma^\tau_{2k}$ with $\tau=X,Y,Z$ and $1 \leq k \in \mathbb{Z}$,
applied first, then applying the even-leading two-spin terms, e.g., $\hat\sigma^\tau_{2k}\hat\sigma^\tau_{2k+1}$, followed by the single-body terms (if any). In the XYZ ordering, the terms involving $\hat\sigma^X$ are applied first, followed by the terms involving $\hat\sigma^Y$, and  finally those involving $\hat\sigma^Z$. It is known that the odd-even ordering introduces less Trotter error than the XYZ ordering~\cite{childsNearlyOptimalLattice2019}. 
Here, we investigate whether the odd-even ordering is also a better choice for simulating the Schwinger model, which includes both nearest-neighbor and (nearly) all-to-all spin-spin interactions.

Since $\hat{H}^x$ in the Hamiltonian in Eq.~(\ref{eq:Hspin}) includes only nearest-neighbor interactions, one way to define an odd-even-ordered product formula is to apply the odd-even ordering to terms in $\hat{H}^x$ only. Defining $\hat{H}^{x}_{n,m}$ to be the term in $\hat{H}^{x}$ acting on sites $n$ and $m$, and, similarly, $\hat{H}^{ZZ}_{n,m}$ to be the term in $\hat{H}^{ZZ}$ acting on sites $n$ and $m$, this ordering can be written as:
\begin{align}
     \hat{\mathcal{S}}^\text{oe1}_{1}(\delta t)&=
     e^{-i\delta t\,  \hat{H}^{Z}}
     e^{-i\delta t\,  \hat{H}^{ZZ}}\ \nonumber\\
    &\times \prod_{k=1}^{(N/2)-1}e^{-i\delta t\, \hat{H}^x_{2k,2k+1}}
    \prod_{k=1}^{N/2}e^{-i\delta t\, \hat{H}^x_{2k-1,2k}}.
    \label{eq:Soe1}
\end{align}
Alternatively, the odd-even ordering can be applied to both $\hat{H}^x$ and the nearest-neighbor terms in $\hat{H}^{ZZ}$, followed by the application of the non-nearest neighbor terms in $\hat{H}^{ZZ}$ as well as $\hat{H}^Z$,
\begin{align}
    \hat{\mathcal{S}}^\text{oe2}_{1}(\delta t)&=e^{-i\delta t\,  \hat{H}^Z} \prod_{n=1}^{N-1}\prod_{m=1}^{n-2}e^{-i\delta t\, \hat{H}^{ZZ}_{m,n}}
    \nonumber\\
    &\times \prod_{k=1}^{(N/2)-1}e^{-i\delta t\, \hat{H}^{ZZ}_{2k,2k+1}}e^{-i\delta t\, \hat{H}^x_{2k,2k+1}}
    \nonumber \\
    &\times
    \prod_{k=1}^{N/2}e^{-i\delta t\, \hat{H}^{ZZ}_{2k-1,2k}}e^{-i\delta t\, \hat{H}^x_{2k-1,2k}}.
\end{align}

The two odd-even-ordered evolution operators are related by a rotation around the $Z$-axis:
\begin{align}
    \hat{\mathcal{S}}_1^{\text{oe1}}=\Bigg(\prod_{k=1}^{N/2}e^{-i\delta t\, \hat{H}^{ZZ}_{2k-1,2k}} \Bigg)
    \hat{\mathcal{S}}_1^{\text{oe2}} 
    \Bigg(\prod_{k=1}^{N/2}e^{i\delta t\, \hat{H}^{ZZ}_{2k-1,2k}} \Bigg).
\end{align}
As a result, if the state is initialized and measured in the $Z$-basis, one arrives at the same measurement outcome using either scheme.
Therefore, in the following we will only consider $\hat{\mathcal{S}}^\text{oe1}_1$ because it requires implementing fewer single-qubit gates.
\begin{figure}[t!]
\centering
\includegraphics[width=0.45\textwidth]{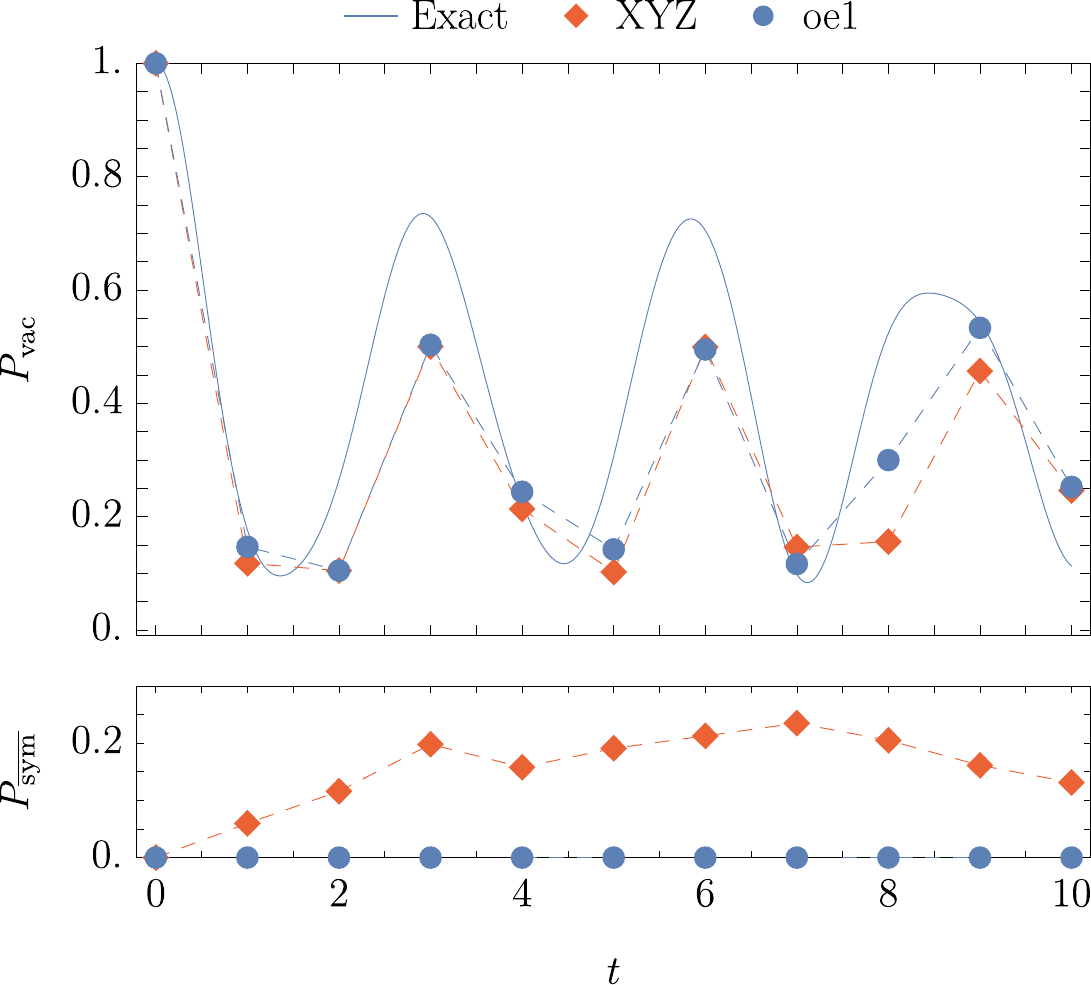}
\caption{Numerical simulation of the projection on the bare-vacuum state $P_\text{vac}$ (upper panel) and the population of symmetry-forbidden states $P_{\overline{\text{sym}}}$ (lower panel) when the system is initialized in the bare-vacuum state for $N=6$, $\mu=0.1$ and $x=0.6$. Different term orderings for the Trotterized evolution are considered: the odd-even ordering defined in Eq.~(\ref{eq:Soe1}) (blue dots) and the XYZ ordering defined in Eq.~(\ref{eq:SXYZ}) (red diamonds). The blue line denotes the exact evolution.}
\label{fig:trotter-order}
\end{figure}

The symmetry operator $\hat S_z$ commutes with $\hat{\sigma}^X_n \hat{\sigma}^X_{n+1}+\hat{\sigma}^Y_n \hat{\sigma}^Y_{n+1}=\hat{\sigma}^+_n \hat{\sigma}^-_{n+1}+\text{H.c.}$, which is not broken up by the odd-even ordering scheme.
Therefore, this ordering does not result in any leakage to the symmetry-forbidden subspace. This is unlike the XYZ ordering that implements
\begin{align}
     \hat{\mathcal{S}}_1^{XYZ}(\delta t)&=
     e^{-i\delta t\,  \hat{H}^{Z}}
     e^{-i\delta t\,  \hat{H}^{ZZ}}
    \nonumber\\
    &\times\prod_{k=1}^{N-1}e^{-i\delta t\, \hat{H}^{(YY)}_{k,k+1}}\prod_{k=1}^{N-1}e^{-i\delta t\, \hat{H}^{(XX)}_{k,k+1}}\ ,
    \label{eq:SXYZ}
\end{align}
where $ \hat{H}^{(XX)}_{k,k+1}$ and $ \hat{H}^{(YY)}_{k,k+1}$ are the terms in the Hamiltonian in Eq.~(\ref{eq:Hspin}) proportional to $ \hat{\sigma}^X_k \hat{\sigma}^X_{k+1}$ and $ \hat{\sigma}^Y_k \hat{\sigma}^Y_{k+1}$, respectively. The size of the leakage to the symmetry-forbidden subspace for both schemes are verified numerically in \cref{fig:trotter-order}.

\subsection{Gate complexity
\label{sec:gate-count}}
There are different approaches to digitally simulating the Schwinger model.
Ref.~\cite{shawQuantumAlgorithmsSimulating2020b} truncates the gauge-boson degrees of freedom in the electric-field basis,  $|E| \leq \Lambda$. The electric-field Hilbert space at each link is then encoded using $\log_2(2\Lambda+1)$ qubits. With OBCs and a zero incoming electric-field flux, the exact theory is recovered for $\Lambda=N/2$. To simulate the model with $N$ sites for time $t$, this approach requires $\mathcal{O}(N+N\log_2N)$ qubits and, up to logarithmic corrections, $\O{N^{5/2}t^{3/2}}$ two-qubit gates, using the second-order Suzuki-Trotter formula. In contrast, our approach integrates out the gauge bosons, leaving only the fermionic degree of freedom associated with the matter fields~\cite{hamerSeriesExpansionsMassive1997,martinezRealtimeDynamicsLattice2016}. Therefore, it requires $\mathcal{O}(N\log_2N)$ fewer qubits to simulate the same model. However, in this approach, one trades the local gauge-matter interactions for long-range matter-matter interactions that correspond to a Coulomb force. Here, we investigate whether this purely fermionic model increases the gate complexity of the simulation compared with the local formulation involving both fermions and bosons.

Given a fixed error tolerance, one can use \cref{eq:general-error-bound} to estimate the required number of Trotter steps in the purely fermionic formulation for simulation time $t$, system size $N$, and at fixed $x$ and $\mu$:
\begin{align}
    \O{\lambda \gamma^{1/p} t^{1+1/p}} = \O{N^{2+1/p}t^{1+1/p}},
 \end{align} 
Here, $\lambda=\O{N^2}$, which is determined by the interactions in $\hat H^{ZZ}$, and $\gamma = \O{N}$, which results from summing over only the interactions in $\hat H^x$.\footnote{We group the interactions in $\hat H^{ZZ}$ and $\hat H^{Z}$ into the $K$th term in the estimation of the Trotter error in \cref{eq:general-error-bound}. Since the terms in $\hat H^{ZZ}$ and $\hat H^{Z}$ commute, further Trotterizing the evolution under $\hat H^{ZZ}$ and $\hat H^{Z}$ into one- and two-qubit gates introduces no additional Trotter error.}
Since each Trotter step requires $\O{N^2}$ two-qubit gates, the gate complexity of the $p$th-order product formula is $\O{N^{4+1/p}t^{1+1/p}}$. For the second-order product formula ($p = 2$), the gate complexity reduces to $N^{9/2}t^{3/2}$, which is a factor of $N^2$ larger than the scaling of Ref.~\cite{shawQuantumAlgorithmsSimulating2020b}.
\begin{figure}[t]
\centering
\includegraphics[width=0.45\textwidth]{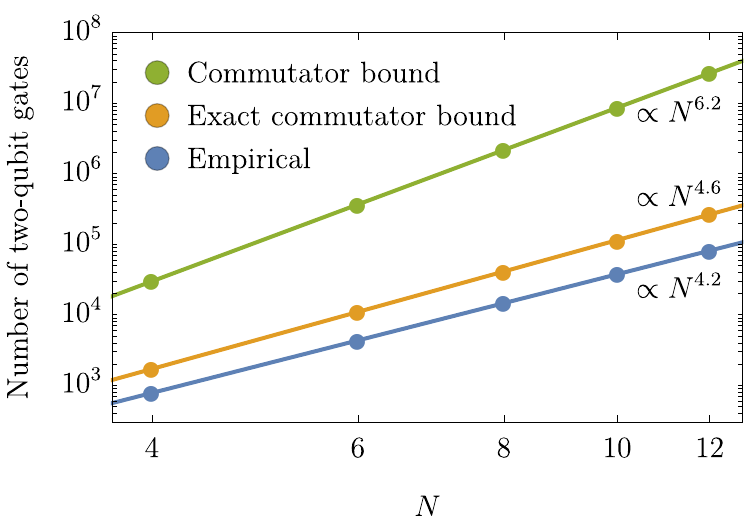} 
\caption{The number of two-qubit gates required to simulate the time evolution under the Hamiltonian in Eq.~(\ref{eq:Hspin}) on an $N$-site lattice for time $t = N$ given an error tolerance $\epsilon = 0.01$, and using the second-order product formula in \cref{eq:2nd-product-formula}.
The commutator bound on the gate complexity (green dots) is estimated from the error bound in \cref{eq:general-error-bound}.
We also obtain a tighter estimate (orange dots) by exactly computing the nested commutators in \cref{eq:exact-comm-bound}.
Finally, the empirical gate count (blue dots) is obtained through a binary search for the minimum number of time steps $t/\delta t$ such that the total error is at most~$\epsilon$.
The straight lines are linear fits that result in the polynomial scaling given in the figure. }
\label{fig:trotter-compare}
\end{figure}

The bound on the Trotter error derived above is an upper bound and the required gate count may be much smaller in practice. In \cref{fig:trotter-compare}, we plot the empirical value of the number of two-qubit gates required for simulating the time-evolution operator in the purely fermionic formulation using the second-order formula for time $t = N$.
The empirical gate count is obtained through a binary search for the minimum number of Trotter steps such that the difference between the Trotterized and the exact evolution is at most $\epsilon = 0.01$. This value is compared to two bounds:
the commutator bound obtained from \cref{eq:general-error-bound}, following the discussion above, and the exact commutator bound derived from computing the norms of the nested commutators in \cref{eq:exact-comm-bound} exactly. The gate complexity $\O{N^{6.2}}$ from the commutator bound agrees with our earlier estimate of $\O{N^{6}}$ for this formulation of the model.
However, \cref{fig:trotter-compare} shows that this bound, obtained by applying the triangle inequality on the norms of the nested commutators, is rather loose. 
Instead, by computing \cref{eq:exact-comm-bound} exactly, hence invoking cancellations between the commutators, the gate complexity reduces to $\O{N^{4.6}}$, very close to the empirical scaling of $\O{N^{4.2}}$.
Notably, the empirical scaling is nearly the same as the scaling of Ref.~\cite{shawQuantumAlgorithmsSimulating2020b} for the fermion-boson formulation.
However, we note that one may also be able to obtain tighter error bounds on the algorithm of Ref.~\cite{shawQuantumAlgorithmsSimulating2020b} empirically.

\subsection{Symmetry protection
\label{sec:sym}}
Digitizing a quantum evolution may introduce errors that populate states forbidden by the symmetry of the target system.
Ref.~\cite{tranFasterDigitalQuantum2020} proposes a method to reduce this leakage when using product formulas and applies it to the fermion-boson formulation of the Schwinger model. In this section, we discuss whether the proposed method is effective at protecting the symmetry of the purely fermionic formulation, and in reducing the Trotter error.

As mentioned earlier, the Schwinger-model Hamiltonian in \cref{eq:Hspin} is invariant under a global rotation around the $Z$ axis of the qubits, i.e., $[\hat H,\hat S_z] = 0$.
Therefore, the expectation value of $\hat S_z$ is conserved if the evolution is exact.
However, due to Trotterization errors, $\langle\hat S_z\rangle$ may deviate from its initial value during the simulation.
To mitigate this error, one can insert rotations generated by the symmetry, i.e., $\hat C(\alpha) = e^{-i\alpha \hat S_z}$, in between the Trotter steps~\cite{tranFasterDigitalQuantum2020}:
\begin{align}
    \hat{\mathcal{U}}(t)=\prod_{k=1}^{t/\delta t} \hat{\mathcal{U}}_k(\delta t) &\longrightarrow \prod_{k=1}^{t/\delta t} \hat{C}^\dag(\alpha_k)\hat{\mathcal{U}}_k(\delta t)\hat{C}(\alpha_k).\label{eq:sym-pro-def}
\end{align}
By choosing suitable angles $\alpha_k$ for different Trotter steps, the errors from each step that do not commute with $\hat S_z$ are rotated by different amounts and can interfere destructively. This results in smaller leakage to the sector with the global charges that differ from the initial state. 
Therefore, the method mitigates errors in symmetry-violating Trotterization schemes, such as the XYZ ordering.
Errors that commute with $\hat S_z$ remain intact under these rotations $\hat C(\alpha_k)$.
As the result, this method has no effect on symmetry-preserving Trotterization schemes, such as the odd-even ordering. Additionally, since the symmetry operator $\hat S_z$ is diagonal in the measurement basis for the observables considered here, one can simply mitigate symmetry-violating errors by post-selecting on measurement outcomes that preserve $\langle\hat S_z\rangle$. We compare the two mitigation methods in the following.

\begin{figure}[t]
\centering
\includegraphics[width=0.45\textwidth]{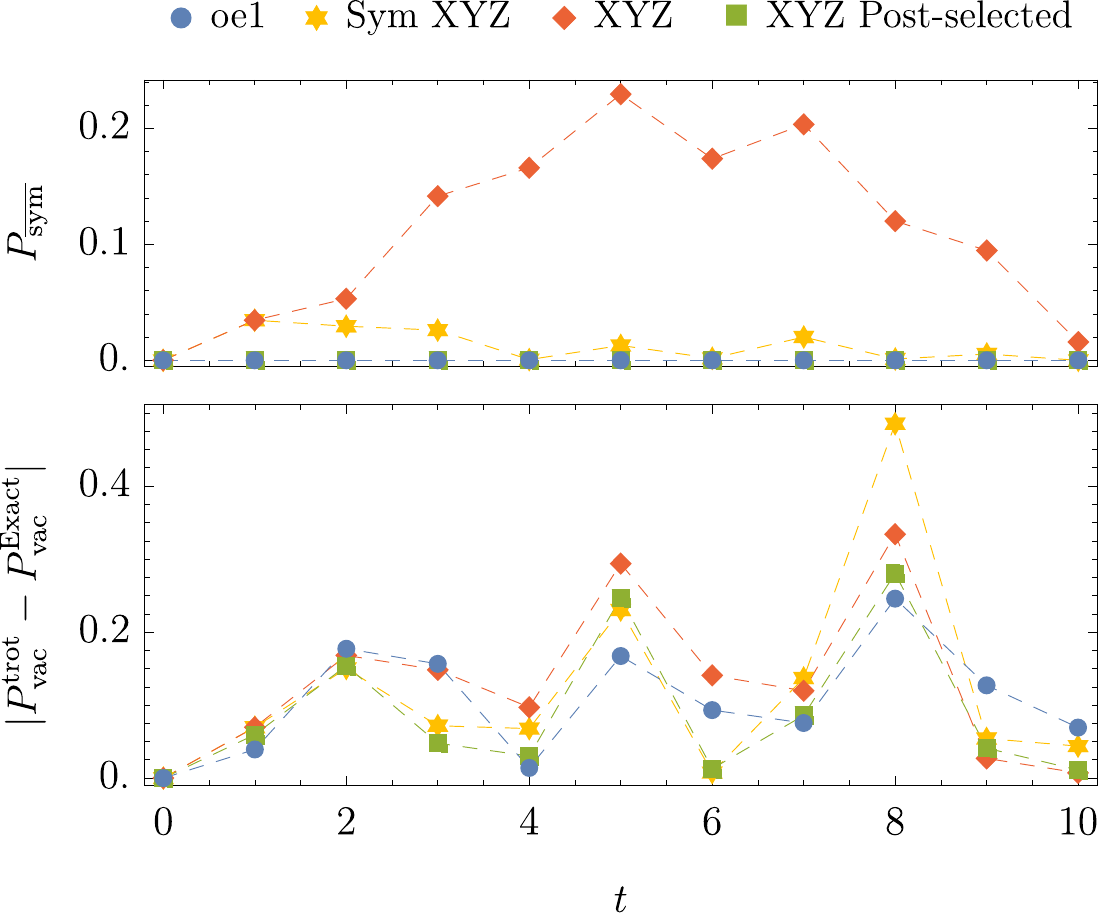}
\caption{The leakage to the symmetry-forbidden sector, $P_{\overline{\text{sym}}}$,  defined as the population in the states with a non-vanishing total charge given a bare-vacuum initial state (upper panel), as well as the error in the bare-vacuum population (lower panel) are shown for the odd-even ordering (blue dots), the XYZ ordering before (red diamonds) and after (green squares) post-selection, as well as the XYZ ordering with symmetry protection but without post-selection (yellow stars). The parameters used for the plot are $\mu=0.1$, $x=0.6$, $N=4$ and $\delta t=1$. The optimal angles for the symmetry-protected simulation are provided in Appendix~\ref{app:alpha}.
}
\label{fig:symmetry-protection-simulation}
\end{figure}

For simplicity, the angles $\alpha_k$ in \cref{eq:sym-pro-def} can be chosen such that $\alpha_k = k\alpha_1$, where for each $t$, $\alpha_1$ is determined by numerically minimizing the leakage to the symmetry-forbidden subspace after $t/\delta t$ steps, as detailed in Appendix~\ref{app:alpha}. As a result, the optimal value of $\alpha_1$ depends on the number of Trotter steps and the simulation time. The top panel in \cref{fig:symmetry-protection-simulation} plots the leakage to the symmetry-forbidden sector for the odd-even ordering, the XYZ ordering before and after post-selection, as well as the XYZ ordering with symmetry protection but without post-selection.
As expected, the odd-even ordering results in no population in the symmetry-forbidden states, and the leakage seen in the XYZ ordering trivially goes to zero after post-selection. Importantly, the symmetry-protected XYZ ordering leads to a strong suppression of the leakage. 

However, this improvement in preserving the symmetry does not guarantee a smaller Trotter error. Given two operators $\mathcal E_\text{sym}$ and $\mathcal E_{\overline{\text{sym}}}$, corresponding to the symmetry-preserving and symmetry-violating errors in the simulation, respectively, it is generally not true that $\norm{\mathcal E_\text{sym}} < \norm{\mathcal E_\text{sym} + \mathcal E_{\overline{\text{sym}}}}$.
In particular, depending on the observable of interest, the effect of the two types of errors may interfere destructively. Therefore, eliminating $\mathcal E_{\overline{\text{sym}}}$ may actually increase the error in the expectation value of the observable. For example, as shown in the lower panel of \cref{fig:symmetry-protection-simulation}, the symmetry protection sometimes increases the error in the population of the bare-vacuum state instead of decreasing it. Overall, symmetry protection does not improve the accuracy of the XYZ-ordering scheme over the odd-even ordering scheme. Notably, post-selection is more successful than the symmetry-protection scheme at mitigating the Trotter error in $P_{\rm{vac}}$.

According to Ref.~\cite{tranFasterDigitalQuantum2020}, the symmetry protection scheme mitigates time-correlated experimental errors as well. Since the errors in our experiment are expected to be dominated by uncorrelated noise, it is interesting to investigate how well the scheme performs in the experimental implementation. To isolate the effects of symmetry protection on experimental rather than Trotter errors, we implement the odd-even ordering, which preserves the symmetry. The results are presented at the end of the next section.
\begin{figure*}[t!]
    \centering
    \includegraphics[width=0.95 \textwidth]{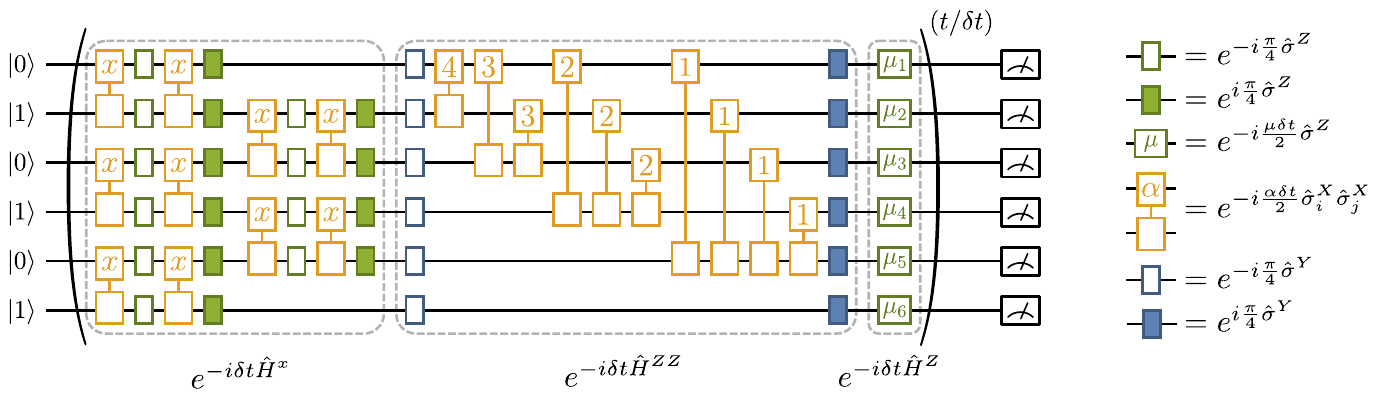}
    \caption{Circuit for  the Trotterized evolution according to the Hamiltonian in Eq.~(\ref{eq:Hspin}) for $N=6$ lattice sites,  with odd-even ordering of the Trotter decomposition introduced in Eq.~(\ref{eq:Soe1}). The interaction term $e^{-i\delta t\hat{H}^x}$ is implemented with nearest-neighbor $X_iX_{i+1}$ and $Y_iY_{i+1}$ gates. The $e^{-i\delta t\hat{H}^{ZZ}}$ term is implemented with $X_iX_{j}$ and $R_i$ rotations. The $e^{-i\delta t\hat{H}^{Z}}$ term involves only $Z_i$ rotations, with the angles $\mu_1=-(\mu+3)\delta t$, $\mu_2=(\mu-2)\delta t$, $\mu_3=-(\mu+2)\delta t$, $\mu_4=(\mu-1)\delta t$, $\mu_5=-(\mu+1)\delta t$, and $\mu_6=\mu\delta t$. Qubits are initialized in the bare-vacuum state $\ket{010101}$, then evolved by repeating the circuit in the parentheses $t/\delta t$ times, and measured individually in the $Z$ basis in the end.}
    \label{fig:circuit6}
\end{figure*}

\section{Experiment and Results}
\label{sec:result}
\noindent
The experiment consists of a chain of up to nine $^{171}\text{Yb}^+$ ions held in a linear Paul trap, with up to six ions used as qubits. The two qubit states are realized in the hyperfine-split ground level, $\ket0 =\ket{^2{S}_{1/2}~ F=0,m_F=0}$ and $\ket1=\ket{^2{S}_{1/2}~ F=1,m_F=0}$. The qubits are initialized to $\ket0$ by optical pumping, and read out on a multi-channel photo-detector using state-dependent fluorescence~\cite{olmschenkManipulationDetectionTrapped2007}. We use two counter-propagating Raman beams to coherently manipulate the states of the qubits~\cite{debnathDemonstrationSmallProgrammable2016}. One Raman beam is split into individual beams to address each qubit separately. The native gates in this setup are single-qubit $Z$ rotations,  $Z_i(\theta)=e^{-i\theta/2~\hat\sigma^Z_i}$, single-qubit rotations around the equatorial plane of the Bloch sphere $R_{i}(\theta,\phi) = e^{-i\theta /2 ~(\hat\sigma^X_i\cos\phi+\hat\sigma^Y_i\sin\phi)}$, and two-qubit gates $X_iX_j(\chi)=e^{-i\chi\hat\sigma^X_i\hat\sigma^X_j}$, where $\phi$ is the angle of the axis of rotation, and $\theta$ and $\chi$ are the rotation angles. The $Z$ rotations are implemented as classical phase advances, and are therefore practically noise free. The single-qubit gates $R_{i}(\theta,\phi)$ are realized by driving a qubit on resonance, with the laser phase set to $\phi$ and the duration proportional to $\theta$. The two-qubit gates use the shared motional modes of the ion chain to mediate interactions between pairs of ions~\cite{sorensenQuantumComputationIons1999,milburnIonTrapQuantum2000,solanoDeterministicBellStates1999a}. Carefully designed amplitude modulation of the laser pulse leaves the qubit states decoupled from the motion at the end of the gate~\cite{choiOptimalQuantumControl2014}. The individual addressing beams allow any pair of ions to be entangled, making the system a fully-connected programmable quantum computer. More details on the setup and gate fidelity are presented in Ref.~\cite{debnathDemonstrationSmallProgrammable2016,landsmanVerifiedQuantumInformation2019,zhuCrossPlatformComparisonArbitrary2021} 

To simulate the evolution under the Hamiltonian in \cref{eq:Hspin}, the product formula needs to be decomposed into the native gates in the experiment.
Specifically, to implement the $Y_iY_j(\chi)=e^{-i\chi\hat\sigma^Y_i\hat\sigma^Y_j}$ ($Z_iZ_j(\chi)=e^{-i\chi\hat\sigma^Z_i\hat\sigma^Z_j}$) gate, each qubit is rotated around the $Z$ ($Y$) axis before and after applying an $X_iX_j(\chi)$ gate. For the odd-even ordering, i.e., $\mathcal{S}^\text{oe1}_1$ given in \cref{eq:Soe1}, the circuit can be broken down into three parts: $e^{-i\delta t\hat H^x}$, $e^{-i\delta t\hat H^{ZZ}}$, and $e^{-i\delta t\hat H^Z}$. An example of the circuit for the Trotterized evolution with $N=6$ is shown in  \cref{fig:circuit6}. The number of gates needed in each Trotter step for the different values of $N$ in the experiment are summarized in \cref{tab:gatecount}. 
While the number of single-qubit rotations can be
further reduced in the circuits, such an optimization is not considered here since they have much lower error rates than the two-qubit gates.

\begin{figure}[h]
\centering
\begin{tabular}[c]{c c}
 &\qquad \fbox{{$N=2$, $\delta t=0.5$}} \vspace*{0.in}\\
\multicolumn{2}{l}{(a)}\\
\multicolumn{2}{c}{
\includegraphics[width=0.45\textwidth]{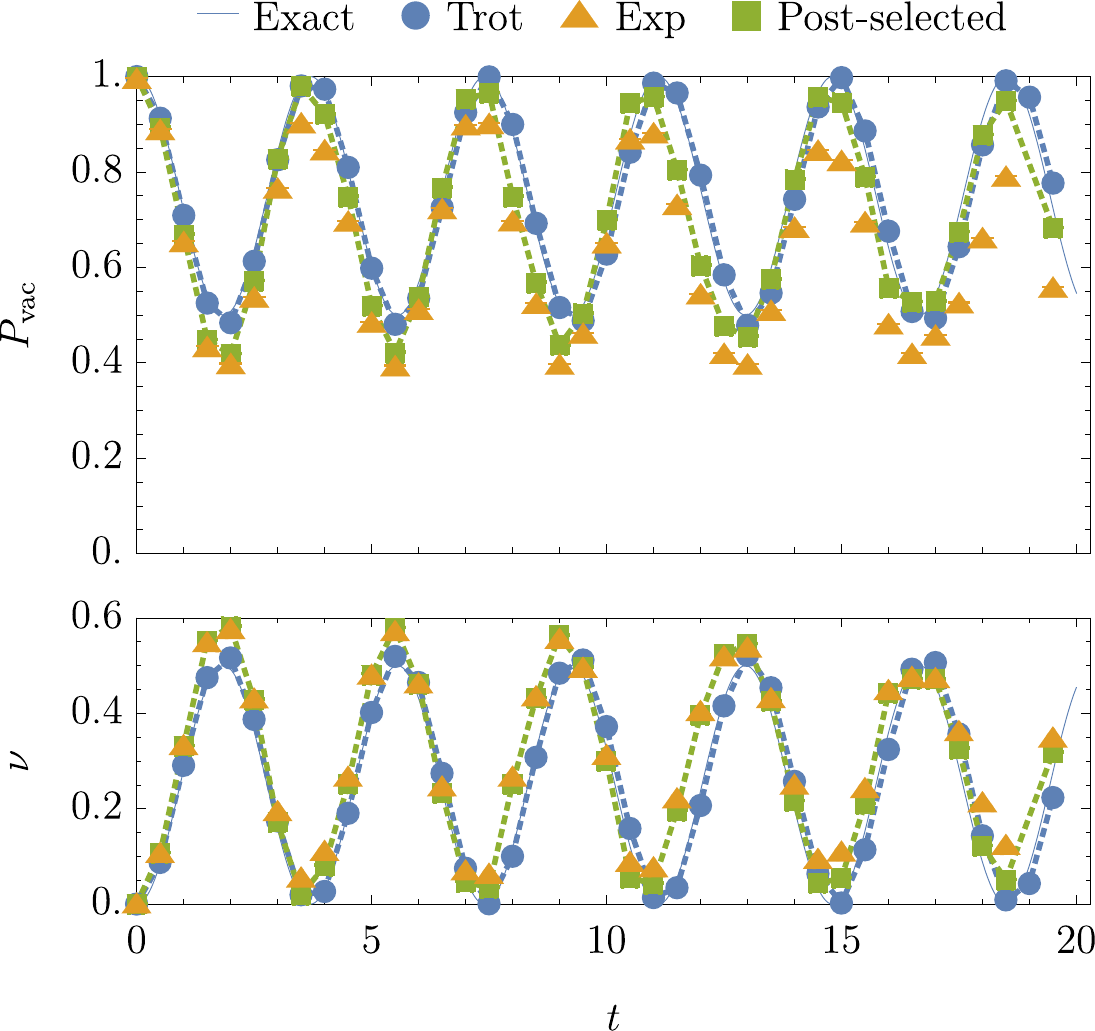}}\\
\multicolumn{2}{l}{(b)}\\
\multicolumn{2}{c}{
\includegraphics[width=0.47\textwidth]{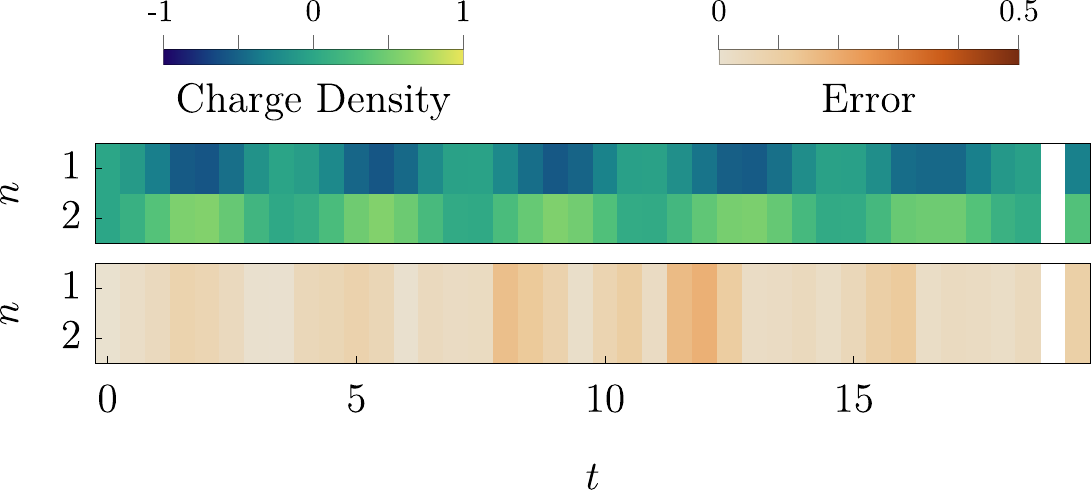}}
\end{tabular}
\caption{Experimental results for $N=2$ and $\delta t=0.5$. (a)~The upper plot shows fluctuation in the bare-vacuum population, $P_{\rm vac}$, while the lower plot shows  particle-number density, $\nu$, as a function of time, indicating the creation and annihilation of the particle-antiparticle pairs.
The dashed lines are a guide to the eye.
(b)~The upper plot shows the local charge density $Q_n$ as measured in the experiment after post-selection, while the lower plot shows its deviation from theory as a function of time.
}
\label{fig:2-qubit-0p5}
\end{figure}
\begin{figure}[h]
\centering
\begin{tabular}[c]{c c}
 &\qquad \fbox{{$N=4$, $\delta t=0.5$}} \vspace*{0.in}\\
\multicolumn{2}{l}{(a)}\\
\multicolumn{2}{c}{
\includegraphics[width=0.45\textwidth]{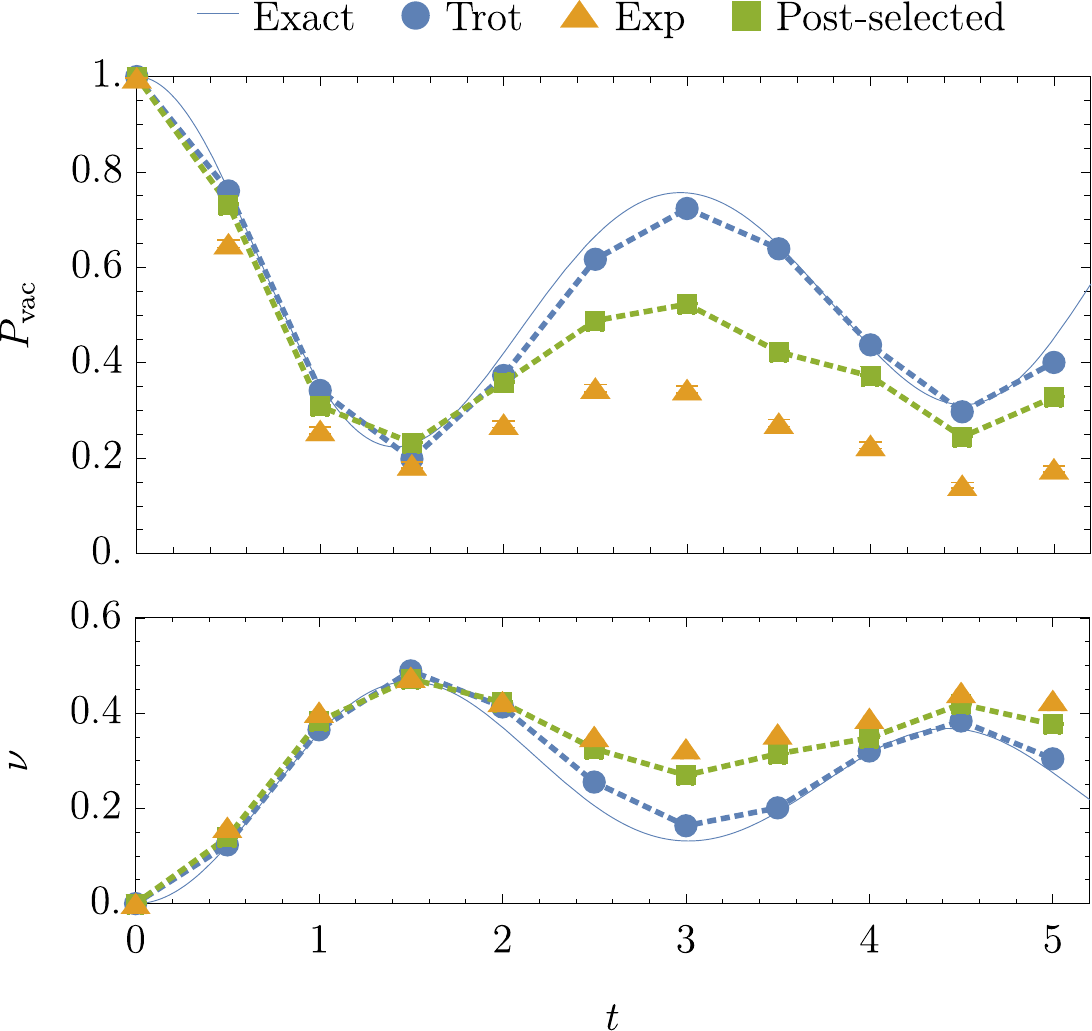}}\\
\multicolumn{2}{l}{(b)}\\
\multicolumn{2}{c}{
\includegraphics[width=0.43\textwidth]{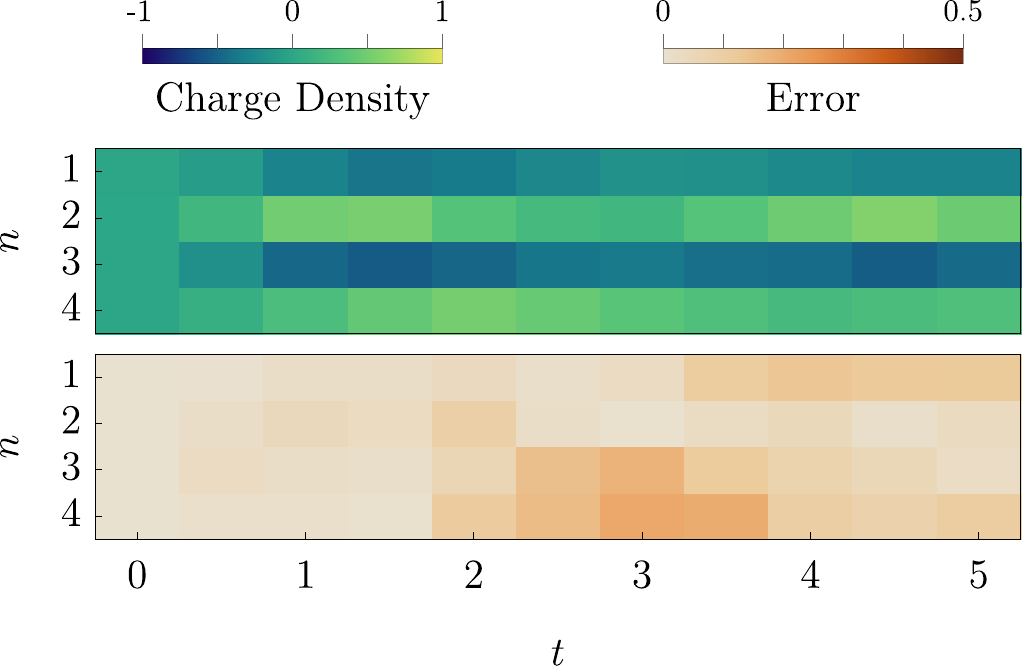}}\\
\end{tabular}
\caption{Experimental results for $N=4$ and $\delta t=0.5$. (a)~The upper plot shows fluctuation in the bare-vacuum population, $P_{\rm vac}(t)$, while the lower plot shows  particle-number density, $\nu(t)$. (b)~The upper plot shows the local charge density $Q_n(t)$ as measured in the experiment after post-selection, while the lower plot shows its deviation from theory.
}
\label{fig:4-qubit-0p5}
\end{figure}
\begin{figure}[t]
\centering
\begin{tabular}[c]{c c}
 &\qquad \fbox{{$N=4$, $\delta t=1$}} \vspace*{0.in}\\
\multicolumn{2}{l}{(a)}\\
\multicolumn{2}{c}{
\includegraphics[width=0.45\textwidth]{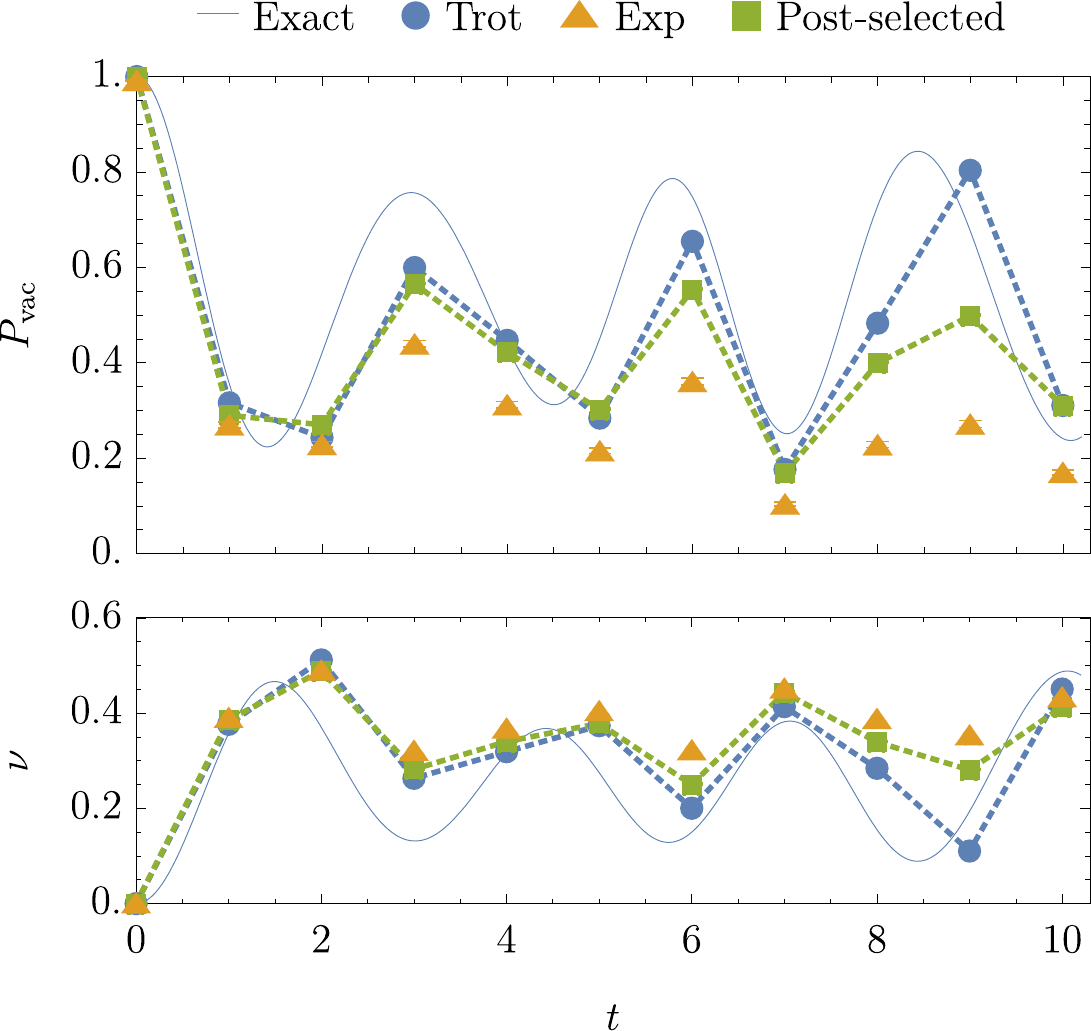}}\\
\multicolumn{2}{l}{(b)}\\
\multicolumn{2}{c}{
\includegraphics[width=0.43\textwidth]{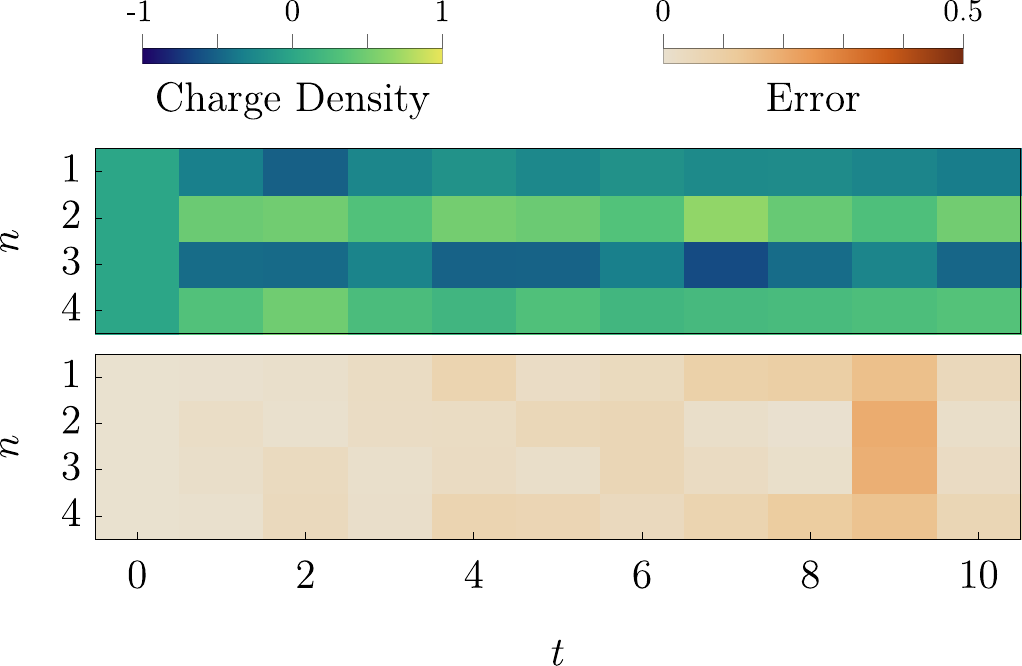}\vspace*{0.in}}
\end{tabular}
\caption{Experimental results for $N=4$ and $\delta t=1$. (a)~The upper plot shows fluctuation in the bare-vacuum population, $P_{\rm vac}(t)$, while the lower plot shows  particle-number density, $\nu(t)$. (b)~The upper plot shows the local charge density $Q_n(t)$ as measured in the experiment after post-selection, while the lower plot shows its deviation from theory.
}
\label{fig:4-qubit-1}
\end{figure}
\begin{figure}[h]
\centering
\begin{tabular}[c]{c c}
 &\qquad \fbox{{$N=6$, $\delta t=1$}} \vspace*{0.in}\\
\multicolumn{2}{l}{(a)}\\
\multicolumn{2}{c}{
\includegraphics[width=0.45\textwidth]{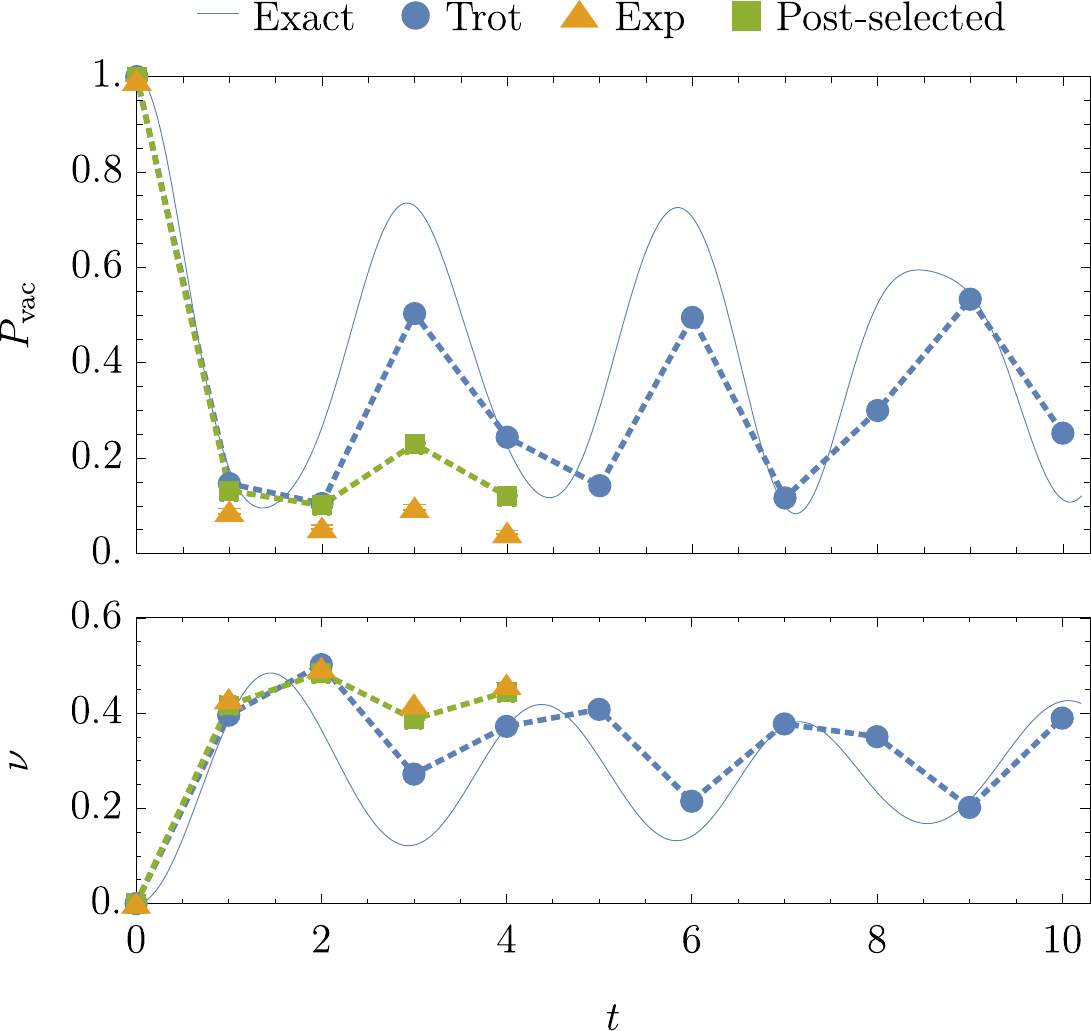}}\\
\multicolumn{2}{l}{(b)}\\
\multicolumn{2}{c}{
\includegraphics[width=0.45\textwidth]{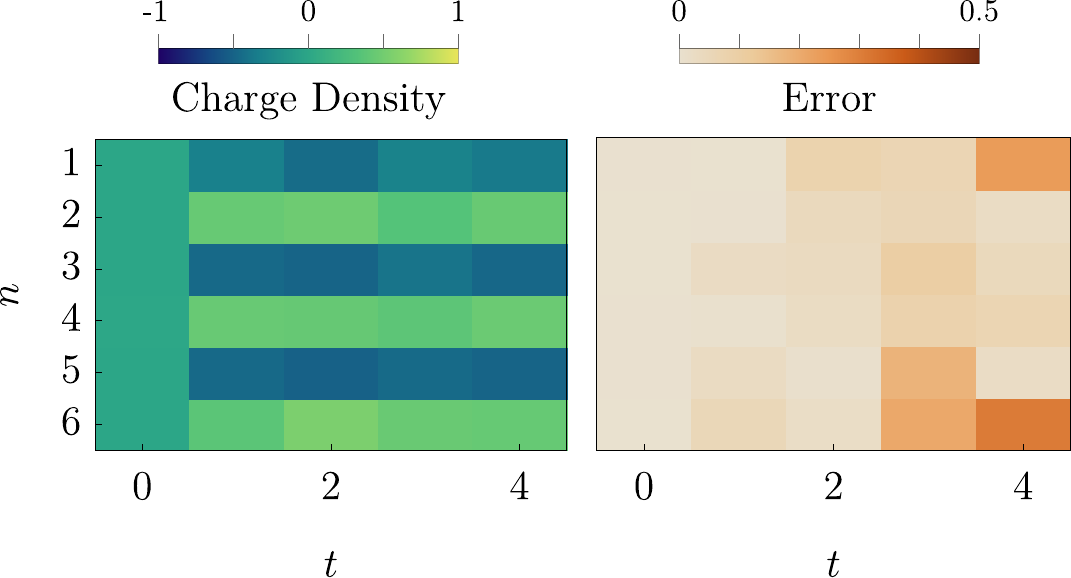}}
\end{tabular}
\caption{Experimental results for $N=6$ and $\delta t=1$. (a)~The upper plot shows fluctuation in the bare-vacuum population, $P_{\rm vac}(t)$, while the lower plot shows  particle-number density, $\nu(t)$. (b)~The left plot shows the local charge density $Q_n(t)$ as measured in the experiment after post-selection, while the right plot shows its deviation from theory. At $t=4$, we reach the gate-depth limit of the hardware.
}
\label{fig:6-qubit-1}
\end{figure}

The experiment starts by preparing the qubits in the vacuum state of the Hamiltonian in the limit of $x=0$, $\ket{\text{vac}}=\ket{\psi_0}$, using $R_i(\pi,0)$ on alternate qubits. The model parameters are set to $\mu=0.1$ and $x=0.6$ to ensure non-trivial dynamics starting from the bare vacuum. 
We measure all the qubits in the $Z$ basis to get the state populations. The state preparation and measurement errors are corrected by applying the inverse of an independently measured state-transfer matrix. The observables of interest are then calculated from the state populations using Eqs.~(\ref{eq:Pvac})-(\ref{eq:Qn}).   

\begin{table}[t!]
    \caption{Gate counts for simulating each Trotter step of the time evolution in the Schwinger model with the odd-even term ordering in Eq.~(\ref{eq:Soe1}), along with the largest number of Trotter steps $t/\delta t$ implemented in the experiment for $N=2,4,6$ staggered sites. 
    }
    \begin{tabularx}{0.45\textwidth}{X Y Y Y Y}
        \toprule
        $N$ & $X_iX_j(\chi)$ & $R_i(\theta,\phi)$ & $Z_i(\theta)$ & $t/\delta t$\\
        \midrule
        $2$ & $2$ & $0$ & $6$ & $39$\\
        $4$ & $9$ & $8$ & $16$ & $10$ \\
        $6$ & $20$ & $12$ & $26$ & $4$\\
        \botrule
    \end{tabularx}
    \label{tab:gatecount}
\end{table}

Algorithmic and experimental errors can, in principle, produce measurement results that break the global charge conservation of the system. This means that starting from the bare-vacuum state, the probability amplitude for evolving to states with non-vanishing total charge may not be negligible. Therefore, to improve the results, one can post-select the measurements so that only outcomes in the relevant symmetry sector are kept. 
Since the odd-even ordering implemented in the experiment does not violate the global charge conservation, these errors result entirely from the experimental imperfections.

\cref{fig:2-qubit-0p5}(a) plots the bare-vacuum population $P_{\rm vac}$,  Eq.~(\ref{eq:Pvac}), and the particle-number density $\nu$, Eq.~(\ref{eq:nu}), as a function of time for $N=2$ and $\delta t=0.5$. For $\delta t=0.5$, the theoretical Trotterized evolution (blue dots) has no significant deviation from the exact evolution (blue line). The  experimental results after post-selection agree well with the theory even after $39$ Trotter steps corresponding to $t=19.5$. Post-selection is shown to significantly mitigate the experimental errors for $P_{\rm vac}$, especially at long evolution times. However, it does not appear as effective for the particle-number density or the survival amplitude of other initial states in general, as plotted in Appendix~\ref{app:population}. \cref{fig:2-qubit-0p5}(b) plots the local charge density $Q_n$, Eq.~(\ref{eq:Qn}), derived from post-selected measurement results and their deviation from the theoretical expectation for the same set of parameters. The local-charge profile is consistent with the global dynamics shown in \cref{fig:2-qubit-0p5}(a) as the pair creation coincides with the destruction of the bare vacuum.

\cref{fig:4-qubit-0p5} plots the same observables as a function of time for $N=4$ and $\delta t=0.5$. Compared with the $N = 2$ case, each Trotter step for $N =4$ requires seven extra two-qubit gates, see \cref{tab:gatecount}, and we only perform ten Trotter steps for this case. Since the experimental error dominates the Trotter error after a few Trotter steps for $\delta t =0.5$, the Trotter step size is increased to $\delta t = 1$ in Fig.~\ref{fig:4-qubit-1}, doubling the simulation time. Even though the Trotter error is larger in Fig.~\ref{fig:4-qubit-1} than in \cref{fig:4-qubit-0p5}, the Trotterized evolution still qualitatively follows the exact solution. In both cases, the experimental data after post-selection agree reasonably well with the numerical simulation.

Next, we increase the number of staggered sites to $N=6$, with the results displayed in \cref{fig:6-qubit-1}. Since each Trotter step now requires $20$ two-qubit gates and $38$ single-qubit gates, only four Trotter steps could be run before decoherence dampens the evolution. Nevertheless, the qualitative behavior, including the first revival of the bare-vacuum amplitude, can still be observed from the results. 

Finally, we study the effect of active symmetry protection by inserting rotations generated by the symmetry operator $\hat S_z$, as discussed in Sec.~\ref{sec:theory}. In the numerical study of Sec.~\ref{sec:theory}, a set of optimal angles was found to minimize the leakage to the symmetry-forbidden subspace caused by algorithmic error. Since such an optimal set cannot be found a~priori, a straightforward strategy is to use a random angle at each Trotter step to average out the leakage after many Trotter steps~\cite{tranFasterDigitalQuantum2020}. Since the effectiveness of the scheme depends on the nature of the experimental error, it is interesting to see if symmetry protection can improve our experimental implementation.

\begin{figure}[t!]
\centering
\includegraphics[width=0.45\textwidth]{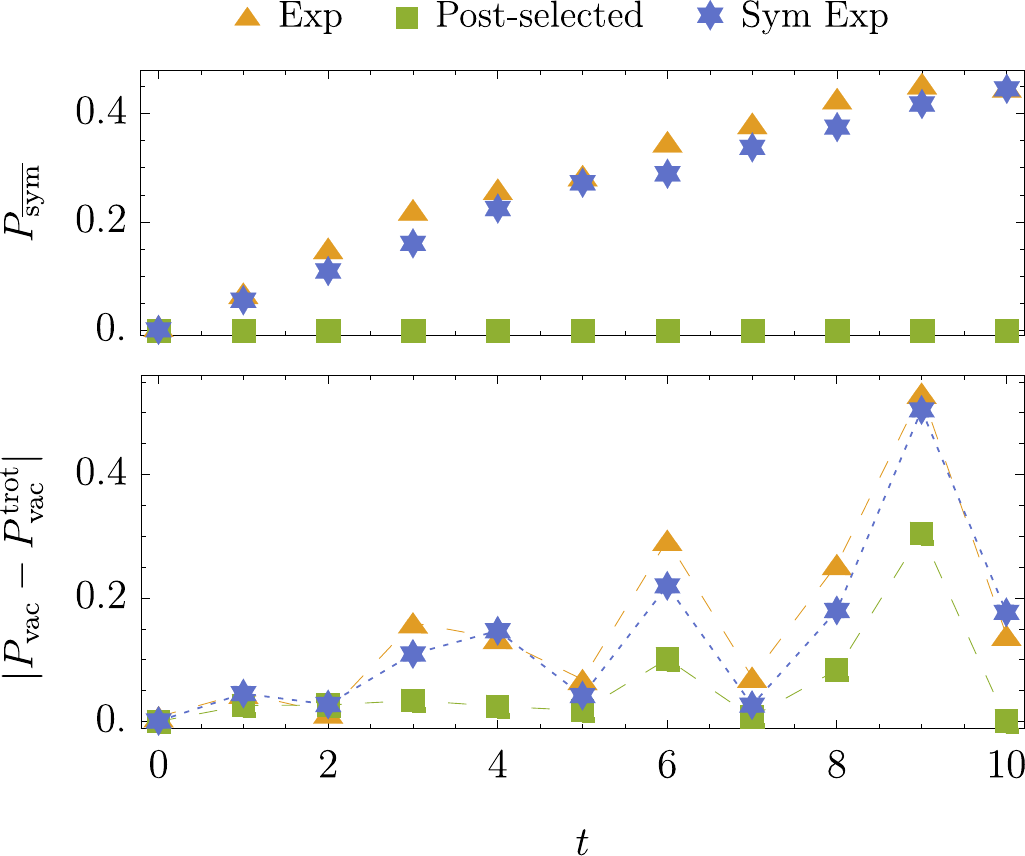}
\caption{Leakage to the symmetry-forbidden subspace (upper panel) and deviation of the experimental results from theory for the bare-vacuum population (lower panel) when different schemes to mitigate errors are applied: no mitigation (orange triangles), post-selection (green squares), and symmetry protection (blue stars).
Note that by definition, the leakage to the symmetry-forbidden subspace is zero after post-selection. In both plots, $N=4$ and $\delta t=1.0$, the system is initiated in the bare-vacuum state, and is evolved via the odd-even ordering scheme.
}  
\label{fig:4-qubit-sym-vs-nosym}
\end{figure}

Figure~\ref{fig:4-qubit-sym-vs-nosym} plots the result of an experiment using the odd-even term ordering. As before, the initial state is the bare vacuum. The unitaries $e^{-i\alpha_k\hat S_z}$, with random angles $\alpha_k$ given in Appendix~\ref{app:alpha}, are inserted between Trotter steps $k$ and $k+1$. While the population in states forbidden by the symmetry, denoted as $P_{\overline{\text{sym}}}$ in the upper panel, decreases with symmetry protection, this reduction is not significant. Furthermore, while the deviation of the bare-vacuum population from the Trotterized theory generally decreases, post-selecting symmetry-preserving measurements appears more effective in mitigating the error in this quantity than the symmetry protection as shown in the lower panel of the figure. This indicates that the experiment is dominated by noise that is not correlated in time. Note that by construction, the symmetry-protection scheme only mitigates time-correlated errors.

\section{Discussion}
\label{sec:discuss}
\noindent
We have digitally simulated the time evolution of the lattice Schwinger model with up to six qubits using the purely fermionic formulation. For a four-qubit simulation, we observe four oscillations of the particle density, and the simulated time is almost four times longer than previously demonstrated using a Trotterized scheme~\cite{martinezRealtimeDynamicsLattice2016,klcoQuantumClassicalComputationSchwinger2018a}. Given the long circuit depths required for dynamical gauge-theory simulations, gate fidelity, rather than qubit number, is the main limitation of our implementation. 
Efforts to overcome such a technical limitation are well underway~\cite{eganFaulttolerantControlErrorcorrected2021}. To mitigate the time-correlated errors, we have applied a symmetry-protection scheme~\cite{tranFasterDigitalQuantum2020} but found negligible effects in suppressing the errors. This symmetry-protection investigation indicates that the dominant noise in experiment is incoherent and uncorrelated.  Incoherent errors can be mitigated by post-selection of the experimental measurements using symmetry considerations. Better-motivated and further-tailored schemes for mitigating incoherent errors are desired in future simulations. Furthermore, it is found that  the symmetry-violating and symmetry-preserving errors can destructively contribute to given quantities, and removing only one of these errors can decrease the overall accuracy. It is therefore important to also develop and apply symmetry-preserving error-suppression schemes in future experiments.

An avenue for improving the quality of the simulation is reducing the gate depth, e.g., by performing gates in parallel instead of sequentially. In our model, $e^{-i\delta t\hat H^x}$, consisting of only nearest-neighbor interactions, can be applied in a fixed circuit depth of $4$ instead of $2N$ by performing all the $X_{2i}X_{2i+1}$ terms, then all the $X_{2i+1}X_{2i+2}$ terms, in parallel. The all-to-all interactions in $e^{-i\delta t\hat H^{ZZ}}$ can be reduced to depth of $N$ instead of $N^2$ if gates $X_iX_{i+n}$, for all $i$ and fixed $n$, are performed in parallel. With trapped ions, parallel operations can be done either in multi-zone architectures~\cite{kielpinskiArchitectureLargescaleIontrap2002a,ryan-andersonRealizationRealtimeFaulttolerant2021}, or in linear chains with advanced control schemes~\cite{figgattParallel2019}.

Alternatively, the gate depth can be reduced by using $M$-body M\o lmer-S\o rensen gates $\text{MS}(\chi,M) \equiv e^{-i\chi \sum_{i=1}^M\sum_{j=i+1}^M\hat\sigma^X_i\hat\sigma^X_j}$ ~\cite{sorensenQuantumComputationIons1999,milburnIonTrapQuantum2000,solanoDeterministicBellStates1999a}. This approach was implemented in Ref.~\cite{martinezRealtimeDynamicsLattice2016} to reduce the number of MS operations in the simulation of the Schwinger model from $\mathcal{O}(N^2)$ to $\mathcal{O}(N)$. In general, a non-trivial optimization of both frequency and amplitude modulation of the beams may be required to implement an M-body gate with the desired rotation angles, as demonstrated in Ref.~\cite{davoudiAnalogQuantumSimulations2020}. Furthermore, one should note that since an $M$-body MS gate has roughly comparable fidelity as that of $M$ 2-body MS ($X_iX_j$) gates~\cite{monz14QubitEntanglementCreation2011}, the overall fidelity of the simulation would likely be similar for both schemes.

When the fermion-boson formulation of the lattice Schwinger model is concerned, a trapped-ion-specific approach to reduce the gate count is to encode the gauge degrees of freedom into the motion of the ions as explained in Ref.~\cite{davoudiSimulatingQuantumField2021}. Besides the standard set of gates, the proposed hybrid digital-analog scheme involves both spin-phonon and phonon-phonon gates. This approach leads to a reduction in both the number of qubits and the number of entangling gates compared with a fully-digital algorithm involving the gauge bosons~\cite{shawQuantumAlgorithmsSimulating2020b}. Future experimental implementations will determine the realistic fidelity of the operations involving dynamical phonons.

To make implementations of more complex gauge theories possible, including non-Abelian and higher-dimensional models,  unifying physics insights, algorithm optimization, hardware implementation, and post-processing is required, as demonstrated in this work. In this context, it would be interesting to investigate whether more resource-efficient encodings of such theories exist, if optimal Trotter decompositions and term ordering schemes can be found, to what degree these preserve local gauge symmetries, whether information regarding the initial state and the symmetries can be incorporated to further tighten the algorithmic error bounds~\cite{su2021nearly, an2021time, yi2021spectral, zhao2021hamiltonian, csahinouglu2021hamiltonian, Rajput:2021khs, tongProvablyAccurateSimulation2021a}, how to balance these errors with experimental errors, and whether symmetry-protection schemes are advantageous in suppressing algorithmic and experimental errors. While progress along these lines is already being made~\cite{klcoSUNonAbelianGauge2019, atas20212, rahman20212, haase2021resource, Raychowdhury:2019iki, davoudiSearchEfficientFormulations2020, Raychowdhury:2018osk, Ciavarella:2021nmj, kan2021lattice, Lamm:2020jwv, Alam:2021uuq, stryker2021shearing, Paulson:2020zjd, Halimeh:2021vzf, Ciavarella:2021lel, Klco:2021lap}, further technological advances in quantum hardware are essential to enable advanced gauge-theory simulations in the upcoming years.

\section*{Acknowledgments}
\noindent
We acknowledge Andrew Shaw’s involvement at early stages of this work. We are grateful to Guido Pagano for valuable discussions. ZD is supported in part by the U.S. Department of Energy's (DOE's) Office of Science Early Career Award, under award no. DE-SC0020271, and by the Maryland Center for Fundamental Physics. ZD further acknowledges the RIKEN Center for Accelerator-Based Sciences, Wako, Japan for their support during earlier stages of this work. ZD and NML acknowledge support by the DOE's Office of Science, Office of Nuclear Physics, under Award no. DE-SC0021143. NML received support from the National Science Foundation QLCI (OMA-2120757) and PFC (PHY-1430094). NML acknowledges
funding by the Maryland-Army-Research-Lab Quantum Partnership (W911NF1920181). AMG is supported by
a Joint Quantum Institute Postdoctoral Fellowship. 
MCT acknowledges the Quantum Algorithms and Machine Learning grant from NTT, number AGMT DTD 9/24/20.

\appendix

\section{Optimal and random angles for symmetry protection
\label{app:alpha}}

\noindent
In \cref{fig:symmetry-protection-simulation}, we study how well the symmetry-protected time evolution given in \cref{eq:sym-pro-def} suppresses the symmetry-breaking Trotter errors for a simulation with the XYZ ordering with $\mu=0.1$, $x=0.6$, $N=4$ and $\delta t=1$. As mentioned in \cref{sec:sym}, for the time evolution with a particular $t$, the angle of rotation, $\alpha_{k}$, for the $k$-th Trotter step depends on $\alpha_{1}$ in the first Trotter step as $\alpha_{k}=k\alpha_{1}$, where $0\leq\alpha_{1}<2\pi$. We choose $\alpha_{1}$ to be the smallest angle with which the leakage to the symmetry-broken subspace at time $t$ is minimized.

\begin{figure}[H]
\centering
\includegraphics[width=0.42\textwidth]{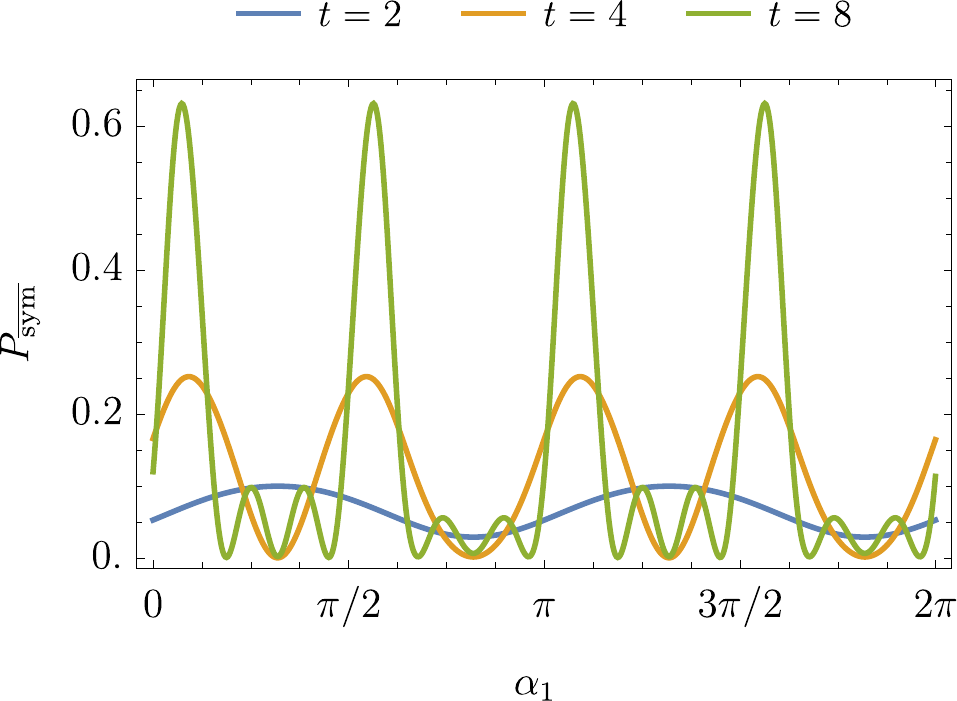}
\caption{The population of the symmetry-forbidden subspace as a function of the angles $\alpha_{1}$ for $t = 2, 4, 8$, for $\mu=0.1$, $x=0.6$, $N=4$ and $\delta t=1$. We choose $\alpha_{1}$ to be the smallest angle that minimizes $P_{\overline{\text{sym}}}$ at each $t$. 
}  
\label{fig:4-qubit-sym-opt-angle}
\end{figure}

\Cref{fig:4-qubit-sym-opt-angle} displays a few examples of how the leakage varies as a function of $\alpha_{1}$ for selected times $t$. As shown in the plot, there are several values of $\alpha_{1}$ that minimize the leakage and the number of minima tends to increase with increasing $t$. The optimal value of $\alpha_{1}$ is determined from the smallest minimum found at each $t$. These values are listed in \cref{tab:angle-sym-XYZ}.

Meanwhile, the angles $\alpha_{k}$ for the symmetry-protected evolution in the experiment in  \cref{fig:4-qubit-sym-vs-nosym} are chosen at random from a uniform distribution in the interval $[0,2\pi]$. These angles are listed in \cref{tab:rand-angle-sym-XYZ}.

\begin{table}[H]
    \centering
    \caption{Values of $\alpha_{1}$ used for the symmetry-protected XYZ ordering in \cref{fig:symmetry-protection-simulation} at different times $t$, together with the corresponding leakage to the symmetry-forbidden subspace.}
    \begin{tabularx}{0.3\textwidth}{Y Y Y}
        \toprule
        $t$ & $\alpha_{1}$ & $P_{\overline{\text{sym}}}$\\
        \midrule
        $1$ & $0$ & $0.0346$\\
        $2$ & $0.8184\pi$ & $0.0294$\\
        $3$ & $0.8184\pi$ & $0.0260$\\
        $4$ & $0.3183\pi$ & $0.0006$\\
        $5$ & $0.8811\pi$ & $0.0128$\\
        $6$ & $0.2314\pi$ & $0.0018$\\
        $7$ & $0.4370\pi$ & $0.0201$\\
        $8$ & $0.1875\pi$ & $0.0011$\\
        $9$ & $0.8496\pi$ & $0.0055$\\
        $10$ & $0.9887\pi$ & $0.0000$\\
        \botrule
    \end{tabularx}
    \label{tab:angle-sym-XYZ}
\end{table}

\begin{table*}[]
    \centering
    \caption{Randomly chosen values of $\alpha_{k}$ used in the experiment to study the effect of symmetry protection on experimental error. The result is presented in \cref{fig:4-qubit-sym-vs-nosym}.}
    \begin{tabularx}{0.9\textwidth}{Y Y Y Y Y Y Y Y Y Y Y}
        \toprule
        $t$ & $\alpha_{1}$ & $\alpha_{2}$ &$\alpha_{3}$ &$\alpha_{4}$ &$\alpha_{5}$ &$\alpha_{6}$ &$\alpha_{7}$ &$\alpha_{8}$ &$\alpha_{9}$ &$\alpha_{10}$ \\
        \midrule
 1 & \text{0.9559$\pi $} & --- & --- & --- & --- & --- & --- & --- & --- & --- \\
 2 & \text{1.1461$\pi $} & \text{0.2987$\pi $} & --- & --- & --- & --- & --- & --- & --- & --- \\
 3 & \text{0.0150$\pi $} & \text{0.6927$\pi $} & \text{1.6279$\pi $} & --- & --- & --- & --- & --- & --- & --- \\
 4 & \text{0.5861$\pi $} & \text{0.0333$\pi $} & \text{0.0787$\pi $} & \text{0.1613$\pi $} & --- & --- & --- & --- & --- & --- \\
 5 & \text{1.7496$\pi $} & \text{1.7986$\pi $} & \text{0.4505$\pi $} & \text{1.4374$\pi $} & \text{0.3222$\pi $} & --- & --- & --- & --- & --- \\
 6 & \text{1.7205$\pi $} & \text{0.0706$\pi $} & \text{1.0666$\pi $} & \text{1.5912$\pi $} & \text{1.0554$\pi $} & \text{0.8444$\pi $} & --- & --- & --- & --- \\
 7 & \text{0.2499$\pi $} & \text{0.2990$\pi $} & \text{0.1212$\pi $} & \text{0.6793$\pi $} & \text{0.9988$\pi $} & \text{0.9218$\pi $} & \text{1.8565$\pi $} & --- & --- & --- \\
 8 & \text{1.7974$\pi $} & \text{0.7531$\pi $} & \text{0.0548$\pi $} & \text{0.4236$\pi $} & \text{1.8081$\pi $} & \text{1.7279$\pi $} & \text{0.4253$\pi $} & \text{0.8807$\pi $} & --- & --- \\
 9 & \text{0.8880$\pi $} & \text{0.3908$\pi $} & \text{1.7202$\pi $} & \text{1.7779$\pi $} & \text{1.1028$\pi $} & \text{1.7425$\pi $} & \text{1.6552$\pi $} & \text{0.0604$\pi $} & \text{0.2346$\pi $} & --- \\
 10 & \text{0.7112$\pi $} & \text{1.1025$\pi $} & \text{1.3913$\pi $} & \text{0.5387$\pi $} & \text{1.7179$\pi $} & \text{1.0585$\pi $} & \text{0.2870$\pi $} & \text{0.8636$\pi $} & \text{1.6639$\pi $} & \text{1.5434$\pi $} \\
        \botrule
    \end{tabularx}
    \label{tab:rand-angle-sym-XYZ}
\end{table*}

\section{State populations
\label{app:population}}
\begin{figure}[H]
\centering
\vspace*{0.2in}
\includegraphics[width=0.45\textwidth]{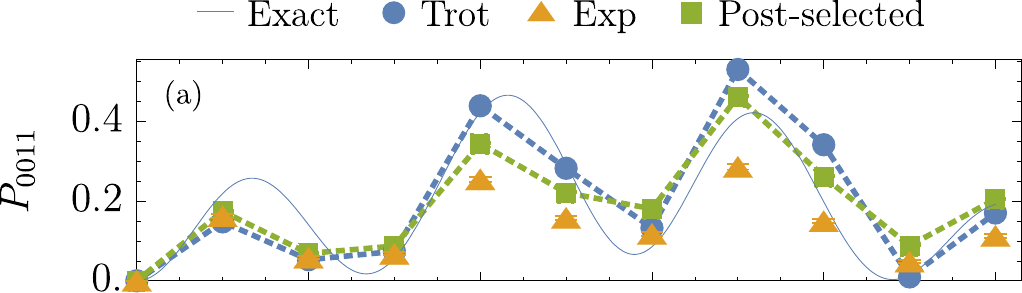}\\
\vspace*{0.05in}
\includegraphics[width=0.45\textwidth]{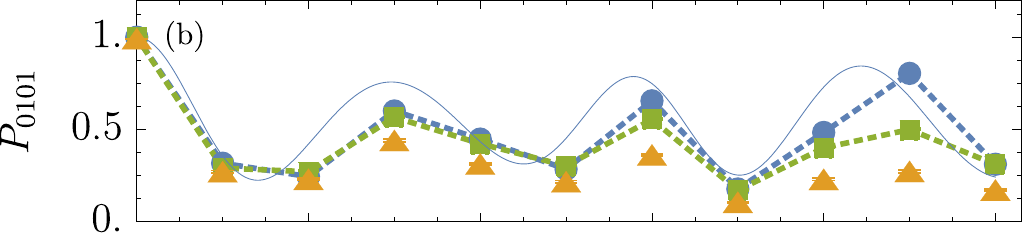}\\
\vspace*{0.05in}
\includegraphics[width=0.45\textwidth]{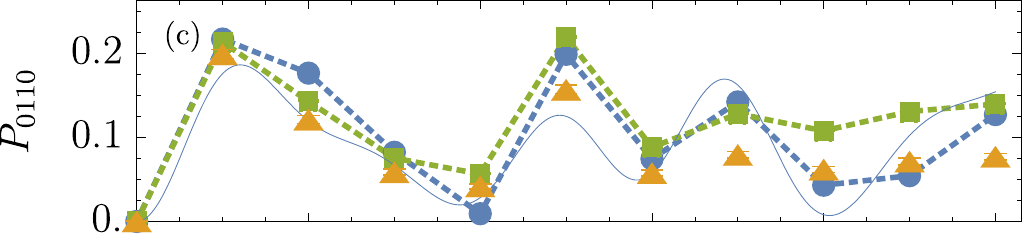}\\
\vspace*{0.05in}
\includegraphics[width=0.45\textwidth]{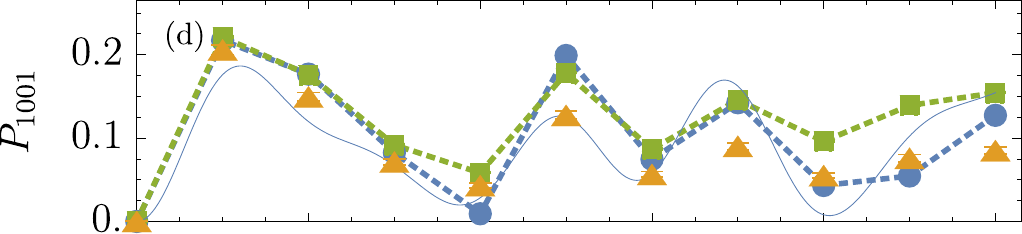}\\
\vspace*{0.05in}
\includegraphics[width=0.45\textwidth]{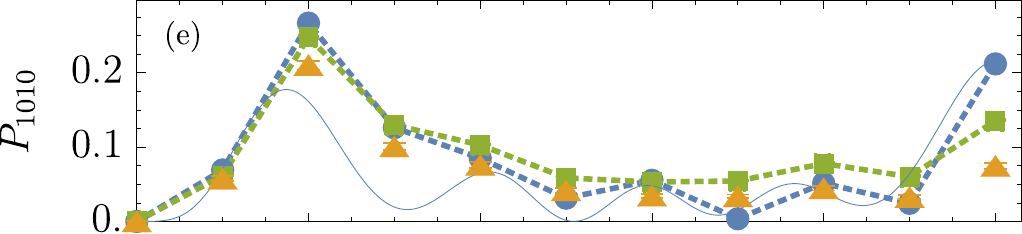}\\
\vspace*{0.05in}
\includegraphics[width=0.45\textwidth]{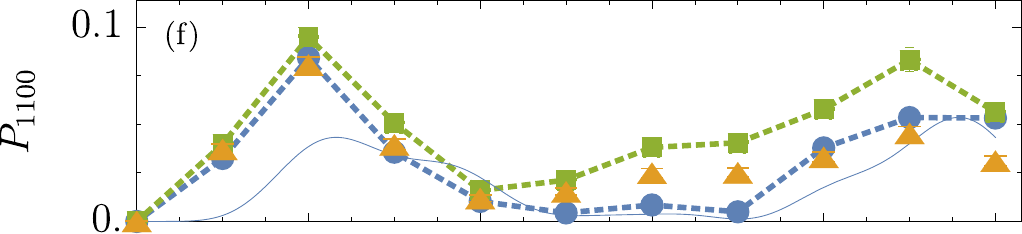}\\
\vspace*{0.05in}
\includegraphics[width=0.45\textwidth]{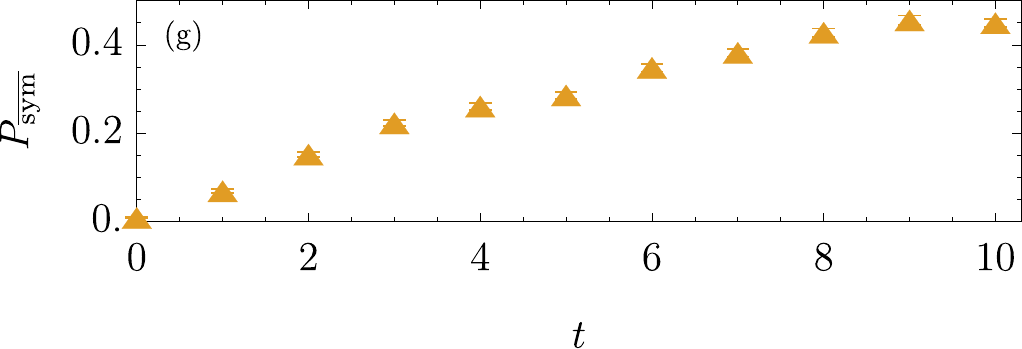}
\caption{Evolution of states characterized by the quantity $P_{\Psi}\equiv |\braket{\Psi|e^{-it\hat{H}}|\psi_0}|^2$ in the symmetry-allowed (a-f) and symmetry-forbidden (g) subspace for $N=4$ and $\delta t=1.0$ starting from the bare-vacuum state $\ket{\psi_0}$. Note that $P_{0101} \equiv P_{\rm vac}$, which is also plotted in Fig.~\ref{fig:4-qubit-1} of the main text. The effectiveness of post-selection in mitigating errors is both quantity and time dependent, and in some cases, it ceases to improve the agreement with theory at larger times.
}
\label{fig:4-qubit-pop-all}
\end{figure}

\noindent
The population in the bare vacuum as a function of time is displayed in \cref{fig:4-qubit-1} of the main text. For completeness, we also plot in \cref{fig:4-qubit-pop-all} the population of all allowed states as a function of time for the case of $N=4$ and $\delta t=1.0$. Additionally, the cumulative population in the symmetry-forbidden sector, $P_{\overline{\text{sym}}}$, is shown, demonstrating the rate at which the leakage to the symmetry-forbidden sector grows.

\bibliography{schwinger}
\end{document}